\documentclass[12pt,preprint]{aastex}
\usepackage{amsmath}
\usepackage{amssymb}
\usepackage{emulateapj5}

\newcounter{qub}
\newcommand{\qq}{\addtocounter{qub}{1}\arabic{qub}}
\newcommand{\kms}{km~s$^{-1}$}

\shorttitle{{LSB Galaxies in the SDSS. I.}}
\shortauthors{Kniazev et al.}

\begin{document}

\title{
Low-Surface-Brightness Galaxies in the Sloan Digital Sky Survey. I.
Search Method and Test Sample}

\author{Alexei Y. Kniazev\altaffilmark{1,2,5}, Eva K. Grebel\altaffilmark{1},
Simon A. Pustilnik\altaffilmark{2,5}, Alexander G. Pramskij\altaffilmark{2,5},
Tamara F. Kniazeva\altaffilmark{2}, Francisco Prada\altaffilmark{1, 3, 4},
and Daniel Harbeck\altaffilmark{1}}

\email{kniazev@mpia.de, grebel@mpia.de, sap@sao.ru, pramsky@sao.ru,\\
tkn@sao.ru, fprada@ing.iac.es, dharbeck@mpia.de}

\altaffiltext{1}{Max-Planck-Institut f\"{u}r Astronomie, K\"{o}nigstuhl 17, D-69117
      Heidelberg, Germany}
\altaffiltext{2}{Special Astrophysical Observatory, Nizhnij Arkhyz,
      Karachai-Circassia, 369167, Russia}
\altaffiltext{3}{Centro Astron\'{o}mico Hispano-Alem\'{a}n, Apartado 511,
       E-04080 Almer\'{\i}a, Spain}
\altaffiltext{4}{Current address: Instituto de Astrof\'{\i}sica de Canarias,
E-38200 Tenerife and The Isaac Newton Group of Telescopes, Apdo 321,
E-38700 La Palma, Spain}
\altaffiltext{5}{Isaac Newton Institute of Chile, SAO Branch}

\begin{abstract}

In this paper we present results of a pilot study to use imaging data
from the Sloan Digital Sky Survey (SDSS) to search for low-surface-brightness
(LSB) galaxies.
For our pilot study we use a test sample of 92 galaxies from the
catalog of Impey et al.\ (1996) distributed over 
93 SDSS fields of the Early Data Release (EDR).
Many galaxies from the test sample are either LSB or dwarf galaxies.
To deal with the SDSS data most effectively a new photometry
software was created, which is described in this paper.
We present the results of the selection algorithms applied
to these 93 EDR fields.
Two galaxies from the Impey et al.\, test sample are very likely artifacts,
as confirmed by follow-up imaging.
With our algorithms, we were able to recover 87 of the 
90 remaining test sample galaxies, 
implying a detection rate of $\sim$96.5\%.
The three missed galaxies fall too close to very bright stars or galaxies.
In addition, 42  new galaxies with parameters similar 
to the test sample objects were found in these EDR fields
(i.e.,  $\sim$47\% additional galaxies).
We present the main photometric parameters of all identified galaxies
and carry out first statistical comparisons.  We tested the quality
of our photometry by comparing the magnitudes
for our test sample galaxies and other bright galaxies with values from
the literature.  All these tests yielded consistent results.
We briefly discuss a few unusual galaxies found in our pilot study,
including an LSB galaxy with a two-component disk and ten new giant
LSB galaxies.
\end{abstract}


\section{Introduction}

Low-surface-brightness (LSB) galaxies are one of the main constituents of
the realm of galaxies.
They are usually defined as objects with a
blue central surface brightness $\mu_{\rm 0}(B)$ significantly fainter than
the Freeman value of 21.65 mag~arcsec$^{-2}$ \citep{Freeman70}.
However, the threshold value of
$\mu_{\rm 0}(B)$ to classify galaxies as LSB galaxies varies in the literature
from $\mu_{\rm 0}(B) \ge$ 23.0 mag~arcsec$^{-2}$
\citep{ImpBot97} to $\mu_{\rm 0}(B) \ge$ 22.0 mag~arcsec$^{-2}$
\citep{ImpBurk01}.
The understanding of the important role of LSB galaxies for many issues of
extragalactic research came during the last 10--15 years
\citep[see, e.g., reviews by][]{Bothun1997, ImpBot97}.
There are many topics for which the knowledge of the properties of
the LSB galaxy population is crucial. They include
the following:
a) the galaxy luminosity function, especially at its faint end,
\citep[e.g.,][]{Dalc98, Trent02, Cross02}, which in
turn is related to the understanding of the primordial power spectrum of
density fluctuations \citep[e.g.,][]{Ostriker93};
b) the spatial distribution of lower-mass galaxies, which allows us to check
the predictions of cold dark matter cosmology for large-scale-structure
formation \citep[e.g.,][]{Peebles01};
c) the physics of star formation at low gas surface densities
\citep[e.g.,][]{vanZee97, Ferg98, Nog01};
d) the role of interactions in galaxy evolution; and many others.

The detection of LSB galaxies
is difficult owing to their intrinsically low global
luminosities and their characteristic low surface brightness.
Despite more than 20 years of LSB galaxy studies,
their census remains incomplete.  
Even in the Local Group new, very faint
dwarf galaxies are still being detected, including four new galaxies within
0.8 Mpc in the past four years
\citep{Armandroff98, Armandroff99, KarKar98, GreGuh99, Whiting99}.
Several samples of LSB objects, identified with different criteria and on
either photographic plates, or on CCD images,
have been published during last 20 years
\citep[for a review see][]{ImpBot97,Dalcanton97}.
They comprise hundreds of galaxies with a central blue surface
brightness, (SB) $\mu_{0}(B)$, in the range of 22 to 26 mag~arcsec$^{-2}$.
The number of known galaxies with lower central SB drops quickly.  In the
range of $\mu_0(B)$~=~24 to 26 mag~arcsec$^{-2}$ only about a hundred or
so are currently known \citep[e.g.,][]{Impey96, Dalcanton97, ONeil97b}.
Very faint LSB galaxies can have
large errors in their photometric parameters or may be artifacts, thus their 
real number is likely smaller.

Past LSB surveys were usually either 
large area photographic surveys or deep CCD surveys with small area
\citep[for a nice review see][]{Dalcanton97}.
The Sloan Digital Sky Survey \citep[SDSS;][]{York2000} is well 
suited for searches for and studies of LSB galaxies due to
its homogeneity, area coverage, and depth (see the SDSS Project Book\footnote{
http://www.astro.princeton.edu/PBOOK/science/galaxies/galaxies.htm}).
The SDSS is an imaging and spectroscopic survey that will eventually
cover about one quarter of the Celestial Sphere.
The imaging data are collected in drift-scan mode
in five bandpasses \citep[$u, \ g, \ r, \ i$, and $z$;][]{SDSS_phot}
using a mosaic CCD camera \citep{Gunn98}.  The SDSS passbands
were carefully chosen to provide
a wide color baseline, to avoid night sky lines and atmospheric OH bands, 
to match passbands of photographic surveys, and to guarantee good 
transformability to existing extragalactic studies.
The SDSS data have already been used in a number of galaxy studies,
for instance in order
to calculate the galaxy luminosity function \citep{Blanton01},
to derive galaxy number counts \citep{Yasuda2001}, to measure the
effect of galaxy-galaxy weak lensing \citep{Fischer00}, to
explore the depth of galaxy potential wells \citep[e.g.,][]{McKay02},
to investigate
statistical properties of bright galaxies \citep{Shimasaku2001}, and to
study the color separation of galaxy types \citep{Strateva2001}.

In this paper we describe the first step of a new LSB survey which was
started with data from the SDSS Early Data Release
\citep[EDR;][]{Stoughton2002}. To develop efficient
methods and to understand possible limitations, in this pilot work we started
with a test sample
from the catalog by \citet{Impey96},
which includes among others a large fraction of LSB galaxies and
luckily covers the equatorial region provided by the SDSS EDR data.
In the current work we present results from a feasibility study
of using the SDSS photometric database
to search for LSB galaxies and to study their basic properties.
Throughout the paper a Hubble constant H$_0$ = 75 km$\,$s$^{-1}$
Mpc$^{-1}$ is adopted.
Furthermore, we investigated the possibility to use 
the reduced SDSS images themselves.
We show the test sample in \S~\ref{txt:test}.
We describe our programs for the search for galaxies with an effective
diameter larger than the predefined value
and for the calculation of
their photometry using SDSS data in \S~\ref{txt:method},
present the results
for our test sample and analyze its SDSS data properties
in \S~\ref{txt:results}, and discuss them
in \S~\ref{txt:disc}.
The conclusions drawn from this study are summarized in
\S~\ref{txt:summ}.

\section{The need for and creation of the test sample}
\label{txt:test}

In order to find the best method for identifying LSB galaxy candidates
in SDSS data, it is helpful to use a test sample with
known properties.
Our test sample is a subsample
from the \citet{Impey96} (ISIB96 hereafter) catalog,
which overlaps with the EDR.
The ISIB96 catalog and hence our test sample
includes  disk galaxies of various types, namely 
both high-surface-brightness (HSB hereafter) and LSB galaxies.
We tried to pick a sufficiently large test sample in order to be
able to study the various factors affecting an effective automatic selection
procedure. The resulting sample consists of 92 galaxies from ISIB96
that lie in 93 ``fields''\footnote{A ``field'' is
SDSS terminology for an SDSS image covering 
$\approx$12\arcmin$\times$10\arcmin.}
of the EDR ``runs''\footnote{A ``run'' is one continuous scan obtained
with  the SDSS imaging camera.}
752 and 756.  Most fields only contain
one ISIB96 galaxy.  Several of these galaxies fall into the
regions near the boundary of 2 adjacent fields,
and thus were detected on both of them.
In a few of the fields several sample galaxies are located.
The whole area of these 93 fields on the sky, accounting for partly
common boundary regions, is about 3 square degrees.
11 objects from this test sample are classified by ISIB96 as dwarf
ellipticals without (dE) or with nuclei (dEn),
while the remaining galaxies comprise various types of disk galaxies ranging
from different types of spirals to Magellanic irregulars (Im) and dwarf
irregulars (dIrr).
In accordance with our intention to use disk galaxies of various types
as a training set, 
the test sample that we selected from ISIB96 covers a range of
central surface brightnesses $\mu_0(B)$ from 18.2 to 26.4 mag~arcsec$^{-2}$.
As follows from our
measurements (see Section~\ref{txt:disc} for details) the fraction
of LSB galaxies is indeed rather high:
$\approx$36\% of the galaxies with $\mu_{\rm 0}(B) \ge$ 23\fm0 arcsec$^{-2}$
or $\approx$70\% of the galaxies with $\mu_{\rm 0}(B) \ge$ 22\fm0 arcsec$^{-2}$.

\section{Detection Algorithm}
\label{txt:method}

LSB galaxies are a subgroup of the general galaxy population
and comprise primarily extended, quiescently evolving disk galaxies, irregular
galaxies, and dwarf galaxies lacking extended starbursts. 
The SDSS pipeline automatically detects point sources and extended sources.
The latter are mainly galaxies and include many LSB galaxies.
Therefore,
we first tested whether it is possible to exploit the SDSS EDR
image database for the efficient selection of extended LSB galaxies and
for the derivation of their main
photometric  and structural parameters.
We carried out these tests on SDSS fields that included galaxies from
our test sample (using the {\tt sdssQT} query tool
\citep[see][for a description and details]{Stoughton2002}) and tried to
recover these known LSB galaxies.
We found that the standard procedures employed for the creation of the EDR
catalogs to classify and  measure photometric
characteristics of objects split many extended galaxies
with sufficiently  prominent knots (such as luminous H\,{\sc ii} regions)
into several separate entities, each listed as a galaxy of its own.
This problem with sufficiently bright and extended galaxies was pointed out
by other authors in the course of their work with EDR catalog data
as well
\citep{Shimasaku2001, Yasuda2001, Stoughton2002, Blanton01}\footnote{%
The situation has improved  with the latest version of the
SDSS photometric
pipeline, PHOTO version 5.4, which will be used for future data releases,
but the shredding problems continue to persist.}
and is known as ``galaxy shredding''.
The latter problem  revealed the need for own photometry software to
recalculate the integrated photometric parameters for the affected galaxies.
Furthermore, many tests with various photometric parameters from the EDR,
in the attempt to figure out the most efficient means to recover the test
sample galaxies, have shown that either a significant (up to 30\%) fraction
of galaxies
were missed, or that with relaxed selection criteria (which improved the
detection rate of the test sample) 
a large number of unwanted objects was found, which in no way are related to
the galaxies we are looking for.  These unwanted objects exceeded
several times the number of galaxies from the test sample, rendering this
search method highly inefficient.  For these reasons, we decided to create
our own detection software.

ISIB96 selected galaxies with 
angular sizes $D \gtrsim$ 30\arcsec, where $D$ is the
major-axis diameter at the limiting isophote of the APM scans
($\mu_{\rm lim}(B)$~=~24.5$\pm$0.5 mag~arcsec$^{-2}$ \citep{Sprayberry96}
or $\mu_{\rm lim}(B) \sim 26$ mag~arcsec$^{-2}$ \citep{Impey96}).
Therefore we have also chosen angular size as the primary criterion to
select candidate galaxies. We used a simple algorithm for the detection
and subsequent 
photometry of galaxies with large angular sizes on the SDSS images.
The new programs that we created for these purposes are based on
the Kitt Peak International Spectral Survey \citep[KISS;][]{Salzer2000}
reduction package \citep{Akn_etal_97, Akn_etal_98} and
some programs from the Astrophysical Institute of Potsdam ({\tt AIP}) package
in MIDAS\footnote{MIDAS stands for Munich Image Data Analysis System,
the data reduction package of the European Southern
Observatory.}
\citep{Lo_etal_93,Vennik96, Vennik00} for adaptive filtering and topological
operations with masks.

The input data for our programs consist of:
(1) -- SDSS direct images ({\tt fpC} files) for
each field in the $u, \ g, \ r, \ i$ and $z$ filters,
(2) -- SDSS tables ({\tt fpObj} files), where all information 
about the objects found in the field with PHOTO \citep{Stoughton2002}
is stored, and
(3) --  tables with astrometric and photometric
coefficients extracted from the SDSS database with the {\tt sdssQT} query 
tool.
We tried to use as much information as possible from the SDSS database to
simplify our software: information about positions of all
objects identified as stars in the SDSS fields, astrometric coefficients,
and coefficients for recalculations of measured fluxes into the SDSS
photometric system.  Knowledge of the positions of stars
is important, since these objects act as contaminants for galaxy
flux determinations.

The processing steps of our programs can be conveniently described
in terms of 8 discrete tasks or modules:
(1) alignment and combining of $gri$ images;
(2) filtering  of the combined image;
(3) object detection;
(4) integrated photometry;
(5) creation of surface brightness profiles (SBP hereafter);
(6) rejection of false detections;
(7) fitting of SBPs;
(8) calculations of total magnitudes.
We briefly outline these tasks below:

\subsection{Alignment and combination of $gri$ images}
\label{txt:method_alig}

The goal of this task is to create a single combined direct image
with better {\it signal-to-noise ratio} ({\it SNR} hereafter) as compared
to the separate $g,r$, and $i$ images.  We concentrate on images obtained
in these three filters since here the greatest photometric depth is
reached.  The combined image is used in the
following steps for the detection of galaxies with large angular sizes.
To work in the astrometric system of the SDSS we selected $r$-frame
coordinates as reference coordinate system. We produce
a geometrical transformation of the {\em u, g, i,} and {\em z} frames 
to the coordinates
of the $r$ frame for each field.
For this we use the coordinates of the 100 bright unsaturated
stars from the SDSS field, which we extract from the {\tt fpObj} table.
Since $g,r$, and $i$ images have different
${\it SNR}_k=S_k/\sigma_k$ (where $k = g,r,i$, and $S_k$ is the 
background level),
and in order to obtain an optimal {\it SNR} ratio \citep{Richter78, Knox98}
we used the weighted factors
\begin{equation}
w_k = \frac{S_k \sigma_g^2}{S_g \sigma_k^2}
\end{equation}
for averaging.  Here the index $g$ denotes the $g$-image, for which we use a
weighting factor $w_g = 1$.
For the estimation of the background level and noise dispersion at this
and all consequent steps we use an
algorithm
that was developed for KISS reductions \citep{Akn97}, which is similar to
that of \citet{Patat03}. This robust algorithm is
based on the calculation
of a histogram distribution for pixel values of the frame
and on fitting a Gaussian function to that histogram distribution.

\subsection{Filtering  of the combined image}

This task serves to filter
the combined image in order to 
decrease the noise of the background and thus to facilitate the detection of
galaxies  with the sufficiently low central  surface brightness.
We consider the problem of filtering as a problem of robust
estimation of an average value in the window where objects of
interest are mixed up with all other sources from the background noise.
Different kinds of non-linear filters were used in the course of various
LSB galaxy searches \citep{Irwin90, Dalcanton97, Armandroff98}.
Proper filtering is a very important step. Our tests showed that
42\% of the galaxies with $\mu_{\rm eff}(r^*) \ge 23$ mag~arcsec$^{-2}$
presented in this paper were detected only after we applied filtering. Without
filtering they were not detected by our programs even when a 
threshold as low as 1.5$\sigma$ of the noise of the combined $gri$ frame
was used.
For $\mu_{\rm eff}(r^*) \ge 24$ mag~arcsec$^{-2}$ the number of
galaxies that remain undetected without filtering is 90\%.

Two criteria were used for the choice of the filter
and smoothing window size:
(1) the requirement to detect all galaxies from the test sample while
    minimizing false detections
and
(2) the requirement that
    the total calculation time required by all our programs  should not exceed
    more than 1--2 months for $\sim$7000 SDSS fields
    that we plan to analyze in future 
    (see Section~\ref{txt:future}).
Finally, after a number of tests, we use the combination of
two filters that  work one after the other with the same window size
of 27 pixels ($\approx$10\farcs5): a smooth-and-clip (SAC) filter
\citep{Sh_Kn_Li_96} and the
fast median smoothing algorithm with circular window,
implemented into MIDAS by \citet{Sher97}.

\subsection{Object detection}
\label{txt:method_detect}

\begin{figure}
    \begin{center}
    \epsscale{1.0}
    \includegraphics[angle=0,width=18cm]{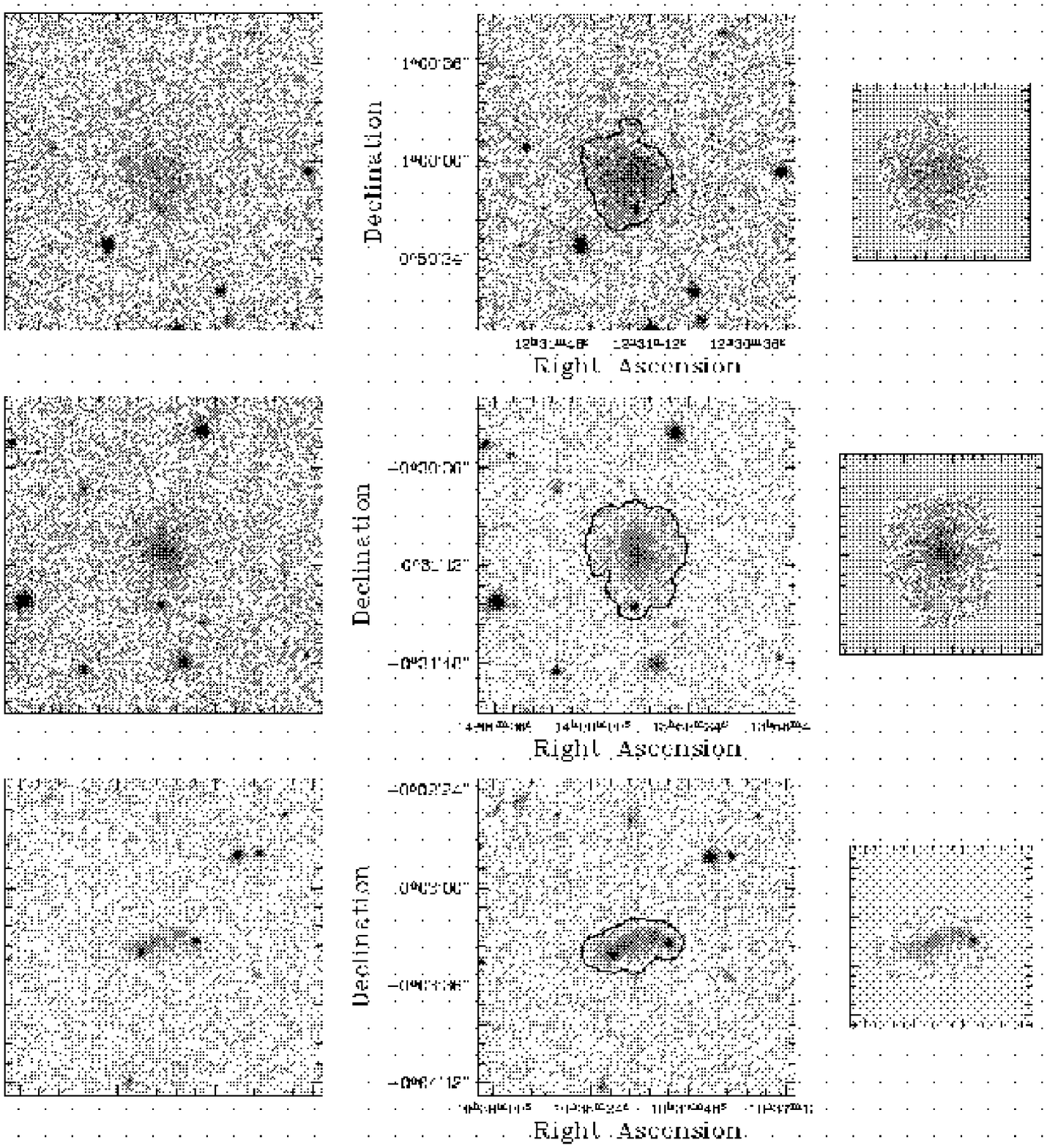}
    \caption{
     Direct images of two galaxies from the ISIB96 test sample,
     SDSS~J123116$+$005939 and SDSS~J135943$-$003133, and of one additional
     galaxy, SDSS~J103806$-$000319, which was detected in the same fields.
     All of these galaxies  were detected as objects with
     R$\ge$12\arcsec\   only after applying the filtering procedure.
     Three images are shown for each galaxy.
     {\it Left panels:} SDSS $r$-band images with the target 
      galaxy in the center,
     {\it Middle panels:} A combined $g, r$, and $i$ image with
     a superimposed isophote corresponding to the threshold level of
     the selection criterion,
     {\it Right panels:} $r$-band extracted subimage, which is then
     stored in
     the temporary work database and prepared for integrated photometry, i.e.,
     with background/foreground sources and sky have been subtracted.
     All images have the same scale, and each object is plotted with the
     same relative low and high intensity threshold values, $-3\sigma$ to 
     $+15\sigma$.
     The images in the left and middle panels have sizes of 
     117\arcsec$\times$117\arcsec. The image sizes in the right panels
     differ from the former since they depend on the mask sizes.
    \label{fig:examples}}
    \end{center}
\end{figure}

We use simple thresholding to detect galaxies with large angular sizes
at the surface brightness level above the limiting isophote.
We  divide this procedure into
two separate iterations:
(1) the detection of objects above the 3$\sigma$ noise level on the
unsmoothed combined images,
(2) additional detection of objects above the 3$\sigma$ noise level on
the smoothed combined images.
In both cases the $\sigma$ of the noise is estimated by the same method
as described in Section \ref{txt:method_alig}.
Mask frames are generated during each step.  These are frames where
topologically isolated  areas have their own unique integer labels.
Such areas are called masks and show the location of the selected objects.
For both steps we are looking for objects with areas of
connected pixels at or above the detection threshold of 2900 pixels.
This corresponds to the area of round objects with the radii
R$\ge$12\arcsec\ or diameters D$\ge$24\arcsec. This radius was selected as
limit to detect the smallest galaxy in our test sample.
In the course of the first iteration we usually detect all bright stars and
bright large galaxies.
Comparison with the positions of the 50 brightest stars from
the {\tt fpObj} table allows us to select all such cases in our list except
for bright stars ($g^*,r^* \le$ 15$^m$)\footnote{%
We note that EDR magnitudes will be referred to with asterisk superscripts
in order to denote their preliminary nature in accordance with the 
recommended SDSS EDR usage \citep[see][]{Stoughton2002}.
As can be seen from \citet[][]{DR1}, the differences between
the $ugriz$ and $u^* g^* r^* i^* z^*$ photometric systems are small,
typically no more than a few hundredths of a magnitude.
},
which were usually classified as galaxies by the SDSS PHOTO software.
We remove bright stars from our list but save their masks
in the mask frame. The two mask frames after two iterations are joined
into one and the areas around the brightest stars are excluded.
Finally, all selected objects have their own mask labels and mask size
with the isophotal level found in the smoothed combined image.
For illustration, several of our detected galaxies with overplotted
mask borders are shown in Figure~\ref{fig:examples}.

No other criteria except the total area are used for the primary selection
step.  Objects with centers that are located closer
than 40 pixels ($\approx$16\arcsec) to the frame borders
are rejected since most of them are detections of ghosts of
bright stars.
The centers of all selected objects are calculated using mask and
filtered frames to exclude effects of background stars and possible
bright H\,{\sc ii} regions. The position angle (PA) and axis ratio are calculated through
the flux-weighted second-moments method \cite[see, e.g.,][]{Stoughton2002}
using masks and combined $gri$ images.

\subsection{Integrated photometry}

This task is intended to derive accurate magnitudes and colors for
the selected objects.
Subimages are extracted from the images in
the $u, \ g, \ r, \ i$ and $z$ bands ($u$ and $z$  images are
transformed to the $r$ coordinate system as described above).
Their centers coincide with the  centers of the selected objects while their
sizes are 1.5 times larger than the masks created in the previous
task.
The latter is necessary for the correct local background determination and
its proper subtraction.
For each object the mask created with the previous task
is increased by 10 pixels ($\sim$4\arcsec) in all directions
before the local background is calculated.
All stars are subtracted from the image using their positions
and Petrosian radii in each filter from the SDSS photometric database.
All previously unrecognized background galaxies and foreground stars
can be masked out by hand with additional circular masks.

A code for fitting the sky background from the {\tt AIP} package
for adaptive filtering is used, which constructs the background within the
masked regions.  This algorithm iteratively fills the
background inside the mask by interpolating the background from the regions
outside the mask \citep{Lo_etal_93, Vennik96, Vennik00}.
This algorithm is used twice: (1) to subtract any contaminating sources
like foreground stars or background galaxies and
(2) to fit and subtract the sky background.
After that the apparent flux is measured inside the same mask
for background-subtracted $ugriz$ images.
The instrumental flux is transformed into the apparent
magnitude in the standard SDSS photometric system.
An example of several of our detected galaxies after preparation for the
apparent magnitude
calculation in the $r$-band is shown in Figure~\ref{fig:examples}.

\begin{figure}
    \begin{center}
    \epsscale{1.0}
    \includegraphics[angle=0,width=18cm]{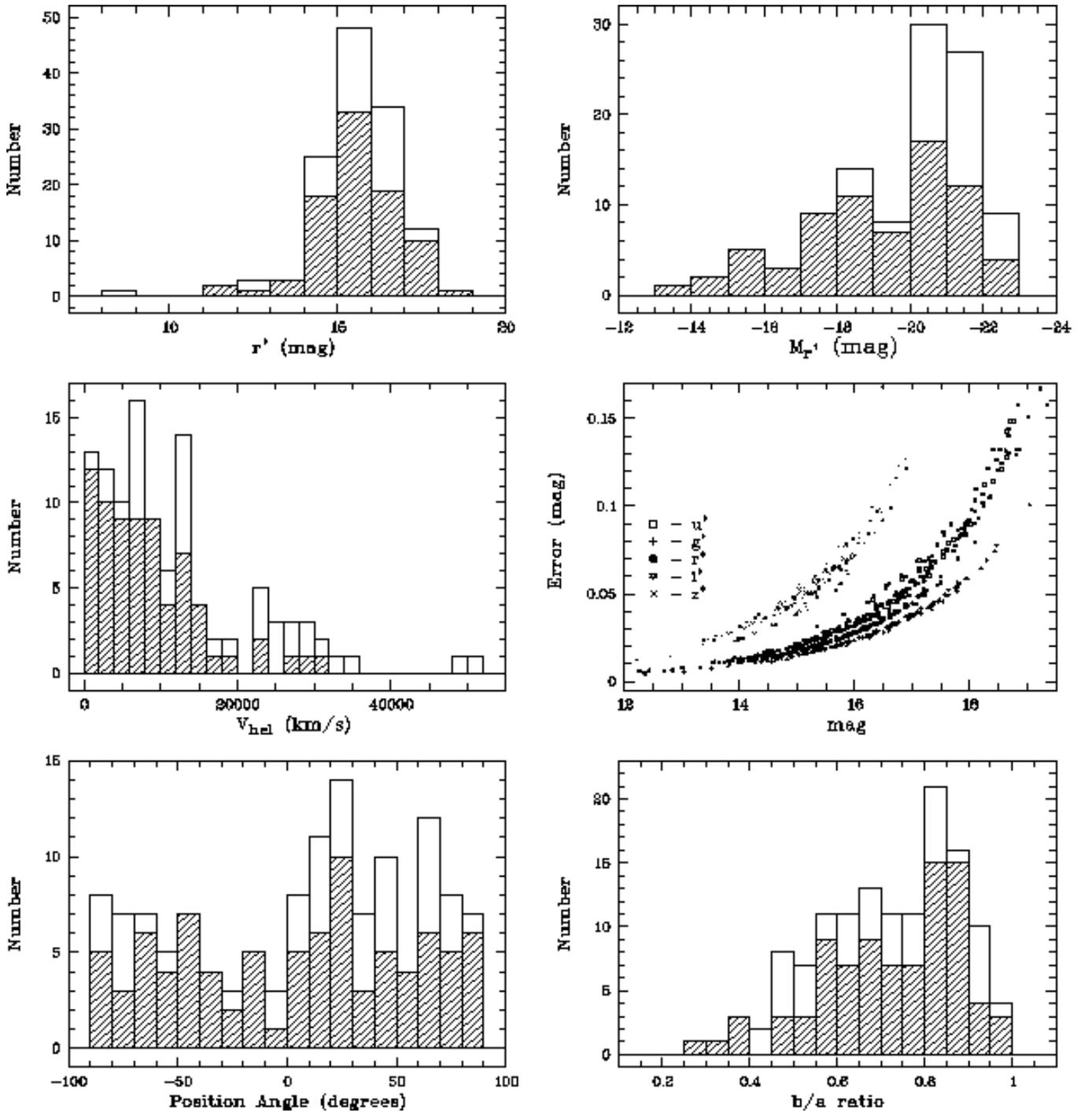}
    \caption{
The distributions of the apparent $r^*$ magnitudes (top-left),
of the absolute M$_{\rm r^*}$ magnitudes uncorrected for Galactic
foreground extinction (top-right),
the heliocentric velocities (middle-left), the
calculated instrumental uncertainties of the integrated magnitudes for
various SDSS filters (middle-right),  the
position angles (bottom-left), and the axis ratios (bottom-right)
for all galaxies detected in the 93 test fields with our programs.
The galaxies from ISIB96 are shown with hashed bins.
    \label{fig5}}
    \end{center}
\end{figure}

The total uncertainty of the instrumental magnitudes includes the
uncertainty of the background determination $\sigma_{sky}$ and
the photon noise $\sigma_{obj}$, which can be calculated
from Poisson statistics.
Therefore the instrumental
magnitude uncertainty $\sigma_{instr}$ is computed for
the cumulative flux $CF$ in the total mask area $N_{pix}$ with the
value of the local background dispersion $SKY$:
\begin{equation}
\label{eq:error}
  \sigma_{instr} = 2.5 \cdot \log{\Big(1 + \frac{\sqrt{\sigma^2_{obj} + \sigma^2_{sky}}}{\sigma^2_{obj}}\Big)},
\end{equation}
where  $ \sigma_{obj} = \sqrt{CF} $ and
$  \sigma_{sky} = \sqrt{N_{pix} \cdot SKY} $.
The distributions of apparent $r$ magnitudes, absolute $r$ magnitudes
(M$_{\rm r}$),
velocities, calculated instrumental errors of the integrated magnitudes,
PAs, and axis ratios for all galaxies found in the 93 test fields
with our programs are shown in Figure~\ref{fig5}.

\subsection{Creation of surface brightness profiles}

The next task creates $u^*g^*r^*i^*z^*$ SBPs of each detected galaxy
using the background-subtracted images, which were used for the calculation of
the integrated magnitudes.
To simplify our programs and to make them more robust with regard
to the creation of
SBPs for galaxies with different morphologies, our software generates SBPs
in the same way as PHOTO does \citep{Stoughton2002}, i.e., 
by measuring magnitudes in circular apertures.
As the analysis in \cite{BBP99} showed, good agreement exists
between an elliptical fit and circular aperture photometry.
Standard SDSS software creates SBPs only for a constant number
of apertures \citep[15 apertures;][]{Stoughton2002} at a fixed spacing.
Our task is capable of measuring SBPs with any constant step size.
We selected 1\farcs0 step sizes as the standard aperture steps for our work.
The uncertainty of each SB level is calculated in the same way as the 
uncertainty of the integrated magnitude (with equation~(\ref{eq:error})).
After a SBP is created, the effective radius R$_{\rm eff}$ (limiting the
region of the galaxy, which contains the half of the light),
the effective surface brightness $\mu_{\rm eff}$ (the mean SB inside
R$_{\rm eff}$), the
radius of the region containing 90\% of the integrated flux (R$_{\rm 90}$),
and the concentration index C=R$_{\rm 90}$/R$_{\rm eff}$ are calculated.
As an example of the output from this task we show in
Figure~\ref{fig:SBPs}
SBPs in $g$ and the radial $(g^*-r^*)$ color profiles for several galaxies from
the test sample.
We also use some programs from the {\tt AIP} package to calculate
the following parameters: the PA of the major axis for each
filter, the axis ratio $b/a$ for each filter,
effective surface brightnesses $\mu_{\rm eff}^{\rm AIP}$,
and effective radii R$_{\rm eff}^{\rm AIP}$.
All these additional parameters are calculated using the multilevel mask
approach \citep{Vennik96, Vennik00}, in which each region of intensities
of the studied object is labelled with a different mask.

Apart from SBPs, our task calculates curves of growth to compare the
final magnitude with the integrated magnitude
calculated with the  previous task. Usually these two
magnitudes practically coincide.
However, for galaxies located near the frame borders, whose masks are
truncated by the borders, our programs in this task calculate
galaxy parameters under the assumption that they are symmetric
relative to their centers.
The best illustration of such a case is NGC~4996, found
as an additional (i.e., non-ISIB96) very extended
galaxy in the fields containing the galaxies from the
test sample. The difference between the magnitude
calculated within the mask area on the frame, and that calculated for the
``restored'', extrapolated image is $\Delta m(g^*) =$~0\fm22.
This galaxy (as it was originally found) is reproduced in the
left panel of Figure~\ref{fig:examples_bilt}. In the right panel of the
same Figure we show its ``restored'' image resulting from the assumption of
symmetry beyond the frame borders.

\subsection{Rejection of false detections}
\label{txt:method_rejection}

After the primary selection algorithm identifies all sufficiently large
objects and their model-independent parameters are calculated,
we need to clean the list of false detections.
This task is performed on an automatic level, subjecting the
detected objects to selection criteria that identify: 
(1) very bright stars not recognized by 
PHOTO as stars and which are therefore selected by our programs in the course
of the object detection procedure;
(2) ghosts close to the borders of the SDSS fields that 
stem from bright stars located outside of our fields;
(3) parts of extended halos around bright stars resembling the
galaxies in which we are interested;
(4) parts of satellite tracks.
All these false detections
selected in the primary selection have either very
unusual colors, or bright effective surface brightnesses
$\mu_{\rm eff}^{\rm AIP}$ calculated with {\tt AIP} package.
To be confident we have checked visually
all the rejected objects in order to estimate the quality of our criteria.
As our statistics show about 80\% of all false detections were rejected
during this stage.

Since our programs are very simple and, by design, are looking
for extended objects, they cannot recognize all possible blends
and other complicated cases (ghosts, parts of halos around bright stars)
if these have photometric parameters
close to ``normal'' galaxies.  Such cases thus cannot be 
rejected by our  task
automatically either. Therefore, during the second step we interactively
checked all remaining candidates by eye to identify such objects.

We applied  our programs to all selected
93 SDSS fields where the 92 galaxies from the test sample are located.
Altogether 245 candidates with sufficiently large angular sizes
($R \ge$12\arcsec) were identified automatically in these
fields in the course of the primary selection procedure.
In the second step, as described above,
most of the remaining bright stars (marked in the
SDSS database as "galaxies" and therefore not rejected during the first step),
their ghosts or parts of halos around them, and linear tracks from satellites
were removed.
This step resulted in the removal of 82 objects
(or $\sim$79\% of all false detections among the non-galaxies).
All objects remaining after these two steps have been checked by eye
to identify possible complicated cases that were not rejected automatically.
This step revealed 22 more false candidates ($\sim$21\% of all false
detections), which were then removed by hand.

\subsection{Fitting of SBPs}

This task produces fits to the $g^*r^*i^*$ SBPs created earlier.
It is well known that the fitting functions should be based
upon the physics of the formation and evolution processes.
Unfortunately, these processes are still not well understood
and the most commonly used functions are derived empirically.
The fitting functions for elliptical galaxies and spiral galaxy
bulges include the King model \citep{King66} and de Vaucouleurs law
\citep{Vauc53}. Recently \cite{APB95} suggested that a generalized version
of the de Vaucouleurs profile ($r^{1/n}$) provides  better bulge fits.
On the other hand, \citet{AS94} and \citet{BBH99} suggest that late-type spirals
often have bulges best fitted by an exponential.
Exponentials \citep{Freeman70} and inner-truncated exponentials
\citep{Korm77} are usually used for the disk components of spiral galaxies.

Our programs follow the scheme to fit both the disk and bulge
components with exponential SB profiles to measure their
central surface brightness $\mu_0$ and scale length $\alpha$.
The equation for such profiles in logarithmic scale usually looks as follows
\citep{Sers68}:
\begin{equation}
\label{equ:exp_profile}
\mu(R) = \mu_0 + 1.086 \cdot (R/\alpha)^n,
\end{equation}
where $R$ is the distance along the axis and $n$ is taken to be 1
for both bulge and disk. In very rare cases for dwarf galaxies
(e.g., \cite{JBF00})
with obvious $n >$~1 we fit the profile with only a disk and with $n$ as
an additional parameter.

This task is interactive and is
based on the MIDAS {\tt FIT} package \citep{MidasA},
which uses different methods for performing a nonlinear least-squares
fit of the data to a specified function.  It also allows control
over the inclusion of different parameters and the range of the data
to be fitted. Our program performs the fitting on the surface intensity
data and accomplishes an optimized fit
by minimizing the weighted rms deviation of the
data from the fit function. It calculates weights as $w_k = \sigma_k^{-1}$,
where $\sigma_k$ is the uncertainty calculated for each SB level
with the previous task for the creation of SBPs.
Some examples of our fitting are presented in Figure~\ref{fig:SBPs}.

We also incorporated the de Vaucouleurs profile
($n$ = 1/4) in our fit task in order to have the  possibility
(1) to automatically classify our observed SBPs by comparing them
to different model light profiles and finding the model profile with the
smallest $\chi^2$ difference, and
(2) to get fit parameters for bright elliptical galaxies in future
applications of our programs.

\subsection{Total magnitudes}
\label{txt:total_mag}

Our method of galaxy detection and the subsequent determination
of their
integrated magnitudes is essentially isophotal photometry with a
sufficiently deep isophotal detection limit ($\mu_{\rm lim}$) that will
be discussed in more detail in Section~\ref{txt:limpar}.
It is well-known that unlike for point sources, the fraction of light
of a galaxy contained within the limiting isophote  $\mu_{\rm lim}$
is a function of
several parameters such as central surface brightness $\mu_{\rm 0}$,
redshift, point-spread function (PSF), and cosmological dimming
(see, e.g., \citet{Dalc98}). In our case
the minimal diameter of the selected galaxies of 24\arcsec\ is much larger
than the PSF, since the typical seeing in the SDSS EDR images
is $\sim 1\farcs5$ \citep{Stoughton2002}.  Thus any sample, selected with
this method from SDSS images, will be independent of the PSF.
In the case of a purely exponential surface brightness profile
(equation (\ref{equ:exp_profile}) with $n=$~1) it can be shown in
simple analytical form that the ratio of the total galaxy flux to that
observed within the limiting  isophote
$\mu_{\rm lim}$ \citep{AllenShu79, Impey88} obeys
\begin{equation}
\label{equ:F_ratio}
\frac{F_{\rm lim}}{F_{\rm tot}} = 1 - (1 + n_\alpha) \cdot e^{-n_\alpha},
\end{equation}
where $n_\alpha$ is the number of scale lengths $\alpha$ observed within
the limiting isophote. And in the general case $n_\alpha$ is given by
\begin{equation}
n_\alpha = \frac{\mu_{\rm lim}-\mu_0 - 10\cdot\log{(1+z)} - k(z)}{1.086},
\end{equation}
where $10\cdot\log{(1+z)}$ is a term accounting for the $(1+z)^4$
cosmological dimming in surface brightness and $k(z)$ is a term to correct
for the redshifting of the spectral energy distribution (the $k$-correction).
The dependence of a fraction of the light above the limiting isophote
on  $(1+z)^4$ is illustrated in \citet{Dalc98}.
In Figure~\ref{fig:ratio_model} we  show the behaviour of the
$F_{\rm lim}/F_{\rm tot}$ ratio versus the central surface brightness
$\mu_{\rm 0}$
for various values of the limiting isophotes.
In real galaxies the situation is, of course, more complicated. 
The true ratio $F_{\rm lim}/F_{\rm tot}$ is higher
for galaxies with detectable bulge subsystems or
S\'ersic SBPs ($n > 1$ in equation (\ref{equ:exp_profile})).

If one wishes to avoid such a systematic bias in the determination
of magnitudes, it is possible to calculate total
magnitudes m$_{\rm tot}$ integrated out to infinity
(for equation (\ref{equ:exp_profile}) with $n = 1$)
using the equation
\begin{equation}
\label{equ:disk_tot}
m_{\rm tot}^d = \mu_0 - 5 \cdot \log_{10} (\alpha) - 2.5 \cdot \log_{10} (2\pi)
\end{equation}
Here $\alpha$ is in arcsec and index `d' refers to a purely exponential
disk. The last term is $\approx$1.995.
It is, however, difficult to get very accurate values for m$_{\rm tot}$,
since they depend on the accuracy of the scale length and $\mu_0$
derived during the fit procedure and thus introduce extra uncertainties
\begin{equation}
\label{equ:disk_err_tot}
\sigma_{\rm model} = \sqrt{\sigma_{\rm \mu_0}^2 + \Big(\frac{5\sigma_{\rm \alpha}}{\alpha}\Big)^2}
\end{equation}
As one can see in Figure~\ref{fig:model_errors} (top and middle panels),
the mean uncertainty of the central surface brightness in, for example, 
the $g$ band is about 0.12 mag~arcsec$^{-2}$,
and the mean relative uncertainty of the
scale length in this filter is  $\sim$4\%.  The respective median
parameters are 0.08 mag~arcsec$^{-2}$ and  $\sim$3\%.
This leads to the mean value of $\sigma_{\rm tot}$ of 0\fm25
(or median value of $\sigma_{\rm tot}$ of 0\fm17),  since
it is $\ge \sigma_{\rm model}$
with equation~(\ref{equ:disk_err_tot}).
To decrease this error and to take any excess non-disk light into
account, our program calculates $m_{\rm tot}$ as the sum of two
components: (1) the integrated magnitude
inside $\mu_{\rm lim}$ (see Section~\ref{txt:limpar} for the determination of
$\mu_{\rm lim}$ values in different SDSS bands)
and (2) a part of an extrapolated disk for $\mu >\mu_{\rm lim}$,
which is easy to calculate from
equation (\ref{equ:F_ratio}) and equation (\ref{equ:disk_tot}).
The full uncertainty in this case is calculated as
\begin{equation}
\label{equ:err_tot}
\sigma_{\rm tot} = \sqrt{\sigma_{\rm instr}^2 + \frac{F_{\rm tot} - F_{\rm lim}}{F_{\rm tot}} \cdot \sigma_{\rm model}^2},
\end{equation}
where $\sigma_{\rm instr}$ is the uncertainty in the instrumental magnitude 
calculated with equation (\ref{eq:error}).
As one can see, for example, $\sigma_{\rm tot}$ for $g$ calculated in
this way (bottom panel of Figure~\ref{fig:model_errors})
is a factor of three lower than the value derived by a standard
method:  the mean value is of 0\fm08 and the median one is of 0\fm06.

In a few cases where we used the S\'ersic profile
(equation (\ref{equ:exp_profile}) with $n > 1$),
to calculate parameters like $F_{\rm tot}$ and
$\sigma_{\rm model}$, we used the modified formula for $m_{\rm tot}$ instead of
formula (\ref{equ:disk_tot}) (see, e.g. \citet{MacArthur02};
our and their $n$ are defined inversely of each other):
\begin{equation}
\label{equ:sers_tot}
m_{\rm tot}^{\rm Sersic} = m_{\rm tot}^d - 2.5\cdot \log_{10} \Big(\frac{1}{n} \cdot \Gamma \Big(\frac{2}{n}\Big)\Big),
\end{equation}
where $m_{\rm tot}^d$ is exactly the expression from formula
(\ref{equ:disk_tot}) and $\Gamma (x)$ is the Gamma function.
For the range of $n$ $(1 \leq n \leq 4)$, which is sufficient for our goals,
it can be well estimated from the approximation of $\Gamma (x)$
by the Stearling series (e.g., \cite{KK61}, formula 24-4-10).
The respective additional term $\sigma_{\rm add}(n)$, depending on $n$,
will appear in the expression for $\sigma_{\rm tot}^2$
\begin{equation}
\label{equ:err_tot_all}
\sigma_{\rm tot} = \sqrt{\sigma_{\rm instr}^2 + \frac{F_{\rm tot} - F_{\rm lim}}{F_{\rm tot}} \cdot (\sigma_{\rm model}^2 + \sigma_{\rm add}^2)}
\end{equation}
It can be estimated with an accuracy of $\sim$10\% as follows:
\begin{equation}
\label{equ:sigma_add}
\sigma_{\rm add} \approx\frac{2.17\cdot\sigma_{\rm n}}{n^2} \cdot \Big(\ln\Big(\frac{2}{n}\Big)+\frac{n}{4}\Big),
\end{equation}
where $\sigma_{\rm n}$ is uncertainty of the index $n$ as one of the
fitting parameters.

\section{Results}
\label{txt:results}

\subsection{Detected galaxies}
\label{txt:Detected}

After the cleaning the list of originally selected objects from
false detections (Section~\ref{txt:method_rejection})
we were left with 141 objects selected as galaxies of sufficiently
large size (with `equivalent' diameter $D \ge$ 24\arcsec).
We also cleaned our sample from the evident
elliptical (E) galaxies, as unwanted non-LSB objects.
They were rejected after the SBP fitting procedure.
Altogether we identified and removed from our sample 9 galaxies
($\sim$6\%) with SBPs typical for ellipticals and
were finally left with 132 galaxies. Of these 132 non-E galaxies 3 objects
were detected independently in two different, partially overlapping SDSS
fields.  Therefore,  we ended up with 129 unique non-E galaxies.

The list of test sample galaxies is given in Table~\ref{tbl-1}.
It also includes additional galaxies that we found in the
same SDSS fields, applying the same selection criteria (see
Section~\ref{txt:Detected}) as for the galaxies from ISIB96 sample.
Therefore, the total number of galaxies in Table~\ref{tbl-1} is 129.
The table contains the following information:   \\
{\it column 1} -- Galaxy name based on the equatorial coordinates for
equinox J2000.\\
{\it column 2} -- Galaxy names as given in the paper by ISIB96.
If not from ISIB96, this entry is blank. \\
{\it columns 3--4} -- Right ascension and declination for equinox J2000.
These coordinates were measured on the SDSS images, using the programs
described in Section~\ref{txt:method}. \\
{\it column 5} -- SDSS image identification (run number, column, and field).\\
{\it column 6} --  Apparent $B$-magnitude and its error. Both were
recalculated from $g^*$ and $r^*$
integrated magnitudes using of equations from
\citet{SDSS_phot1}.\footnote{We did not find any large and/or systematic
differences when using the transformation equations of \citet{SDSS_phot}
or of  \citet{SDSS_phot1} for the calculation of $UBVR_cI_c$ magnitudes
from $u^*g^*r^*i^*z^*$ magnitudes for our galaxies.
We thus adopt the equations from \citet{SDSS_phot1} throughout the paper.}\\
{\it column 7} --  $(B-V)$ color, calculated from $g^*$ and $r^*$ magnitudes
using the respective equations from \citet{SDSS_phot1}.\\
{\it column 8} -- Absolute $B$-magnitude, calculated from the apparent
$B$-magnitude in the previous column and the distance derived
from the known heliocentric velocities, assuming a Hubble flow with H$_{0}$=
75~\kms~Mpc$^{-1}$, with no corrections for the
motion of the Sun and Local Group.
No corrections have been made for Galactic foreground extinction.\\
{\it column 9} -- Heliocentric velocities taken from ISIB96,
from the SDSS EDR database and the
NASA/IPAC Extragalactic Database (NED).\\
{\it column 10} -- Morphological classification in the system of
\citet{Vauc91}. This classification was based on that suggested in
ISIB96 and NED, where applicable.
However, accounting for the superb quality of the SDSS images relative
to most of the previously available data, we have checked the classification
for all sample galaxies and revised it in a number of cases.
For the remaining galaxies the classification was also done by the authors. \\
{\it column 11} -- One or more alternative names, according to the
information from the NED.

Of the 132 detected objects, 87 are from our test sample of 92 galaxies.
They appeared as 90 detections in 93 fields, thus three of
these 87 galaxies were detected twice in overlapping fields.
This implies a detection rate of our algorithms of 87/92, or $\sim$94\%.
In fact, we have found that the two dimmest objects of the ISIB96 test sample
(0956$-$0034 with $\mu_0(B)$=25.8 mag~arcsec$^{-2}$ and
1304$+$0054 with $\mu_0(B)$=26.5 mag~arcsec$^{-2}$ according to ISIB96)
are very likely artifacts, as explained below.
Indeed it is hard to understand how ISIB96 could
detect an object with $\mu_0(B)$=26.5 mag~arcsec$^{-2}$, if their
limiting isophote in the APM scans was $\sim$~26 mag~arcsec$^{-2}$).
Following the analysis of the possible range for $\mu_0$ in
Section~\ref{txt:limpar}, we know that our software is capable
of detecting objects with $\mu_0(g^*) = 26.2$ mag~arcsec$^{-2}$
(recalculated  from $\mu_0(B)=$~26.5 for 1304$+$0054 mag~arcsec$^{-2}$
assuming that the object is very blue with $(g^*-r^*) =$~0\fm3),
if their $\alpha >$~30\arcsec.
For redder objects the detectability is even higher.
But even if these galaxies are less extended than the scale length limit
that we imposed, and thus cannot
be detected by our programs with their fixed limiting parameters,
they should be  visible on the SDSS images.
We have checked carefully
their positions on the combined $g,r,i$ SDSS images, which certainly
are deeper than the UKST plates used by ISIB96. Nothing is seen
at the positions where the ISIB96 finding charts show some fuzzy
objects. 

We also obtained images at the positions of these two galaxies
in the framework of the Key Project for SDSS Follow-up Observations 
at the Calar Alto Observatory in Spain \citep{Gre01}.
We used the Calar Alto Focal Reducer and
Faint Object Spectrograph (CAFOS) at the 2.2-m telescope in 
direct imaging mode.  CAFOS was equipped with
a $2048\times2048$ pixels SITe CCD with an image scale of $0.53''$ 
pixel$^{-1}$, resulting in a field of view of $16' \times 16'$.
The observations consisted of 3$\times$20-min exposures
in the Gunn $g$-band (i.e.,  deeper than the 54s exposures
in the SDSS $g$-band obtained with the SDSS 2.5-m telescope at Apache
Point Observatory).
We did not find anything obvious at the positions of these two galaxies
in these new direct images.  We therefore conclude that the detections
noted by ISIB96 are probably caused by photographic plate defects.
The occurrence of artifacts among some of the faintest galaxies in the
sample of ISIB96 was also noticed by \citet{KarKar98}.
Accounting for these false ``galaxies'' from the test sample,
the real detection rate of our algorithms approaches 87/90 ($\sim$96.5\%).

The three remaining galaxies from the test sample were not detected by our
automatic algorithms  for similar reasons.
1331$-$0002 (name from ISIB96) is near a very bright star.
1025$-$0040 is also near a very bright star, coinciding with  one of
the diffraction spikes of this star on the SDSS images.
1103$+$0010  is located in the halo of galaxy NGC3521 with $B$=10\fm0.
Thus, summarizing the results for the non-detected galaxies from the
test sample,
we conclude that with our rather simple automatic algorithms, we
lose a few galaxies in the vicinity of bright objects. However, their
fraction is fairly small, $\sim$3\% from the test sample.

42 galaxies were found in addition to the 87 detected galaxies from
the test sample within the same 93 SDSS fields. All these objects were selected
objectively with the same criteria that we predefined to recover all galaxies
from the test sample.  To the 90 true galaxies from the test sample,
the additionally detected 42 new galaxies add another 47\%, i.e.,
almost half as many new galaxies in addition to the number previously known.

Several model-independent observational parameters derived from the
integrated photometry of the test sample and of the additionally found
galaxies are presented in Table~\ref{tbl-3}.
It includes the following columns:
$u^*,g^*,r^*,i^*,z^*$ integrated magnitudes with their errors,
axis ratio $b/a$, position angle PA, and some additional parameters
of these galaxies in the $r$-band: the effective radius $R_{\rm eff}(r^*)$,
mean surface brightness inside this radius $\mu_{\rm eff}(r^*)$,
concentration index C($r$), and $a_{26.2}(r^*)$ -- size
for $\mu_{\rm lim}(r^*)$ = 26.2 mag arcsec$^{-2}$.
For all sample galaxies, whose SBPs were fitted with double exponential
disks (disk and bulge), we present in Table~\ref{tbl-4}
the following model  photometric parameters:
$\mu_0^d(g^*)$, $\alpha^d(g^*)$, $\mu_0^b(g^*)$, $\alpha^b(g^*)$,
$\mu_0^d(r^*)$, $\alpha^d(r^*)$, $\mu_0^b(r^*)$, $\alpha^b(r^*)$,
$\mu_0^d(i^*)$, $\alpha^d(i^*)$, $\mu_0^b(i^*)$, $\alpha^b(i^*)$,
where the index `d' refers to a disk and `b' refers to a bulge.
For those sample galaxies, whose SBPs were fitted with the S\'ersic
profile, the respective model parameters
$\mu_0(g^*)$, $\alpha(g^*)$, $n(g^*)$,
$\mu_0(r^*)$, $\alpha(r^*)$, $n(r^*)$,
$\mu_0(i^*)$, $\alpha(i^*)$, $n(i^*)$
are presented in Table~\ref{tbl-5}.
Four obviously interacting systems,
1113$+$0107, 1208$+$0120, 1250$-$0009, and 1327$-$0020,
were not included in the latter two tables.

\subsection{Total versus apparent magnitudes}

In Figure~\ref{fig:appar_total} we show 
the difference between apparent and total $g^*$ and $r^*$ magnitudes versus
apparent magnitudes for galaxies from our sample.
The total magnitudes were calculated as  described in
Section~\ref{txt:total_mag}.
The mean difference $\Delta g$(Apparent$-$Total) is 0\fm04$\pm$0\fm04,
which is well consistent with the mean value of the total error for this
band, $\sigma_{\rm tot}(g^*)$~=~0\fm08. Only 17 galaxies ($\sim$~13\% of
the total number of the sample galaxies) have $\Delta g^* >$~0\fm08, and
only 4 galaxies ($\sim$~3\%) have $\Delta g^* > 2\cdot\sigma_{\rm tot}(g^*)$.
The situation is similar for the $r$ and $i$ filters.
As was described in Section~\ref{txt:method},
the total magnitudes in our method are available only after fitting SBPs.
This is due to the need to extrapolate the apparent light profile.
While developing our software, we tried to use the extrapolation of
curves of growth in order to calculate total magnitudes. However,
we found that in any case we need to choose different types of extrapolation.
They are different for pure exponential disk, S\'ersic profile, 
and bulge plus disk systems, and thus need to be taken into account
when fitting the SBPs.
Therefore, our total magnitudes are derived only for the $g$, $r$ and $i$ filters,
since we confine our SBP determinations to these three filters.
For the $u$ and $z$ filters we do not perform this procedure, since the
uncertainty of the additional light contributions affecting these 
extrapolations is in
most cases significantly larger than the value of the respective term.
As Figure~\ref{fig:appar_total} shows, our apparent magnitudes are often 
a sufficiently good approximation of the total magnitudes.
For example, in Section \ref{txt:phot_comp} we compare our magnitudes and NED
magnitudes for objects with brightnesses up to $B = 15\fm5$
(see Figure~\ref{fig:phot_comp}),
i.e., to $g^* \sim 15\fm0$ for the typical color index $(g^*-r^*) = 0\fm5$.
As one can see in Figure~\ref{fig:appar_total}, most of our galaxies
in this magnitude range have $\Delta g^*$ and $\Delta r^*$ less than 0\fm03.
Therefore it will be only a minor difference when using apparent magnitudes
instead of total magnitudes, if total magnitudes are not available.

\subsection{Photometric quality}
\label{txt:phot_comp}

To examine the photometric quality of our data and to check our
programs we tested our results in two ways:
(1) we compared the magnitudes obtained for the same objects detected in
different SDSS fields (observed either within the same run, or in
two different runs) as an internal consistency check, and
(2) we compared our magnitudes for previously known objects with
the results of published photometry from the literature as an external
consistency check.

Three galaxies from the test sample each were found in 2 different
fields. Thus, they were measured independently twice, and their
integrated magnitudes are
suitable to check the quality of our photometry and reliability of our error
estimates. This
comparison is shown in Table~\ref{tbl:repeat}.
One can see that the integrated
magnitudes for the two independent measurements of the galaxies
SDSS~J113457$-$004514 and SDSS~J144856$-$004337, found in different fields
of the same
run, are consistent in all bands, since they differ only by
1--3$\sigma$ of the cited uncertainties.
This implies that galaxies located in adjacent fields of the same run
do not show significant differences in their total magnitudes.
This repeatability underlines both the data quality and correct functioning
of our programs.
In both fields we are dealing with the same object from the same run. 
Because the adjacent
SDSS fields have  common regions with a width of 128 pixels
\citep{Stoughton2002},
our programs found  these galaxies in two fields, and  the only
difference is global and local background and noise determination.
The same situation holds for  galaxy SDSS~J135230$+$002504,
found in two fields of {\em different} runs.
Therefore, we can state that we do not see any differences
in the calibration between the SDSS runs 752 and 756.
Since one galaxy is too little for statistics, the planned future extension
of this study to a much larger number of SDSS fields
(of the same two interleaving runs, see Section~\ref{txt:future})
will yield many more pairs of independent photometric measurements.

In Figure~\ref{fig:phot_comp} (top-left panel)  we show the comparison of our
integrated
magnitudes
\citep[transformed to $B$-band magnitudes using the equations
from][]{SDSS_phot1}
with $B_{\rm tot}$ magnitudes of ISIB96.
The comparison of our integrated $B$-band magnitudes with those from ISIB96
shows that although the majority of the galaxies exhibit photometric
scatter consistent with the uncertainties for the ISIB96 sample
(typical uncertainties cited by ISIB96 are $\sim$0\fm07 for galaxies
with CCD photometry and $\sigma \sim$0\fm25 for those without),
there are about a dozen galaxies that significantly
deviate from the expected mean difference $\Delta B$(Our$-$Impey)=0.
The latter galaxies result in a weighted r.m.s. of 0\fm55.
We have checked carefully every object exhibiting a
large difference. However, no
reasonable explanation was found. We did not find any new $B$-band photometry
for this sample in \citet{BIS01} or \citet{ImpBurk01}, where instead
the same magnitudes as in ISIB96 are used.

As an additional test, we compared our values, \citep[transformed to
$UBVR_cI_c$-band magnitudes using][]{SDSS_phot1}
and the ISIB96 $B$-band magnitudes  with
an independent dataset, extracted for several bright galaxies from NED.
17 galaxies from our sample also have $B$-band magnitudes listed in the NED,
and 12 of these are galaxies from
ISIB96, which are shown by filled circles on the top-right panel of
Figure~\ref{fig:phot_comp}.
Therefore, even if we consider that NED data come from many different 
sources and do not constitute a uniform
dataset as such, we expect that the comparison of our photometry with NED
photometry and the comparison of ISIB96 photometry and NED data should
show similar trends (if any).

The comparison
shows that our magnitudes (in $B$-band, as well as in other broad bands),
after accounting for the uncertainties cited in NED
(Figure~\ref{fig:phot_comp}, both bottom panels),
 are well consistent with the magnitudes from NED.
In particular, for the 12 ISIB96 galaxies in common with our sample and with
NED the weighted r.m.s. of  $\Delta B$(Our$-$NED) is 0\fm29.
For all 17 galaxies in common with our list and NED the  weighted r.m.s. of
$\Delta B$(Our$-$NED) is 0\fm25.
For four galaxies, which we find to have $B$ fainter
by 0\fm5--0\fm6 than NED, the difference is most likely
due to projected stars, which were subtracted in our data, but presumably
remained in the photoelectric diaphragm photometry, cited by NED from
the Third Reference Catalogue of Bright Galaxies (RC3) \citep{Vauc91}.
The role of superimposed stars in diaphragm photometry is, on average, larger
for fainter galaxies. If we account for this trend in the comparison between
our and
NED photometry, the resulting r.m.s. for $\Delta B$(Our$-$NED) will decrease
to $\sim$0\fm18, fully consistent with the cited NED errors, since our
uncertainties for this range of magnitudes are significantly smaller
(0\fm01 -- 0\fm03).

The comparison of the $B$-magnitudes of galaxies from ISIB96
with NED shows larger scatter,
significantly exceeding the amount expected from their cited uncertainties.
The weighted r.m.s for $\Delta B$(ISIB96$-$NED) is 0\fm8.
For extreme cases the difference between ISIB96 magnitudes and those
listed in NED amounts to up to 1\fm0--1\fm5.
This suggests that besides the random errors in the ISIB96 sample,
some of their galaxies suffer from rather large systematic shifts.
Further support of this conclusion can be found in 
\citet{Sprayberry95},  where one additional measurement is
presented for the galaxy 1226+0105 from ISIB96.
Its magnitude in ISIB96, $B$=15\fm7, is very different from our
value for this galaxy, $B$=16\fm34. \citet{Sprayberry95} obtained CCD
photometry of this galaxy and measured $B$=16\fm10$\pm$0\fm07.
The latter value is already corrected for Galactic extinction,
k-correction, and cosmological dimming $(1+z)^2$.
After the de-correction for dimming
(V$_{\rm hel}$ = 23655 km s$^{-1}$, \citep{Sprayberry95}) and extinction
we get $B$=16\fm27, which is within the 1$\sigma$ uncertainty of
our $B$-magnitude.

As the last test we show CCD $BV$ total photometry from \citet{Salzer89}
for 5 galaxies in common with our sample. We analyzed these measurements
separately from the NED because they are a uniform set of observations
and have fainter magnitudes than the galaxies listed in the NED.
Four of these galaxies are also part of the ISIB96 data.  The comparison 
of these published magnitudes with our photometry is shown
in the right panels of Figure~\ref{fig:phot_comp} with asterisks.
These comparisons show the same picture as above:
(1) the mean difference $\Delta B$(Our$-$Salzer) is $-0\fm06\pm0\fm03$ and
the mean difference $\Delta B$(ISIB96$-$Salzer) is $-0\fm32\pm0\fm07$;
(2) the weighted r.m.s. for $\Delta B$(Our$-$Salzer) is 0\fm06 and
weighted r.m.s. for $\Delta B$(ISIB96$-$Salzer) is 0\fm15
(in both cases only the four common galaxies were used).   This lends
further support to our
conclusion that the absolute accuracy of our SDSS photometry
as compared to external data is $\le 0\fm1$.

Additionally, Figure~\ref{fig:phot_comp} (both bottom panels)
demonstrates that
our software allows us to get reliable magnitudes even for
galaxies as bright as $\sim$10\fm0 in $B$-band from SDSS data, whereas
galaxies this luminous are excluded from the public SDSS photometric database.
Most of the NED magnitudes for passbands other than $B$ 
are from CCD observations, while $B$-band magnitudes are mainly photoelectric.
In the other bands we therefore see even better consistency between
the magnitudes compiled in the NED and our total magnitudes.
Since, unfortunately, there are only a few  measurements
in each band, we estimated the scatter of the combined data in
the remaining UVRI bands. The mean difference $\Delta UVRI$(Our$-$NED)
is 0\fm03$\pm$0\fm04,
and the weighted r.m.s. $\Delta UVRI$(Our$-$NED) is 0\fm12.

Thus, summarizing the results of the comparison of our new photometry with the
$B_{\rm tot}$ magnitudes from ISIB96 and NED we conclude that:
(1) In general the agreement is very good for the different bands,
    and our photometry results are reliable;
(2) The external quality of photometry for SDSS data is about
the same for the different bands and has a precision of $\le 0\fm1$;
(3)  With our software we are capable of obtaining accurate photometry
  also for all bright
  (non-LSB) galaxies within the studied sky region, while photometry
  for many of these galaxies is not available from the SDSS database
  or has larger uncertainties.
  This is an important by-product of our programs,
  since a large fraction
  of bright galaxies has either only 
  photoelectric photometry \citep{Vauc91}
  or CCD photometry only in one or two bands.

\subsection{Quality of structural parameters}

To investigate the quality of structural parameters of our data we
compared them with the data from ISIB96.
We emphasize that our definition of $\mu_{\rm eff}$ is different from
the one used by ISIB96.
Their $\mu_{\rm eff}$ corresponds to the surface brightness in
the narrow annulus with a mean radius equal to $R_{\rm eff}$.
If these parameters would be defined for the same band, then for a pure
exponential disk $\mu_{\rm eff}$ as defined by ISIB96 should be 0\fm70
fainter than according to our definition.

In Figure~\ref{fig:struct_comp} we compare our parameters
R$_{\rm eff}(g^*)$, $\mu_{\rm eff}(B)$, and $\mu_{\rm 0}(B)$ with
the corresponding ones in ISIB96.
Interacting galaxies as complex, composite systems were not included in this
comparison.
As shown in the upper panel of Figure~\ref{fig:struct_comp}, the values
of the model-independent parameter R$_{\rm eff}$ obtained in the present
work for the test sample galaxies are in good agreement for most of
the galaxies in  ISIB96. The difference is within 1--2 combined $\sigma$ of
R$_{\rm eff}$. To estimate this parameter we used $\sigma_{R_{\rm eff}}$
derived directly from our measurements and the estimate for a purely exponential
disk for ISIB96 data, based on their $\sigma_{\rm tot} =$ 0\fm25. This
results in a relative error of R$_{\rm eff} \sim$ 0.15.
Only for five galaxies (that is, 6\%) are the differences significant. They
exceed 2$\sigma$ of the combined error, and it is difficult to explain
them by random errors.
Since all our data are of uniform quality,
we can suggest that a few galaxies of ISIB96 are based on
lower quality data (as follows also from our photometry comparison
in Section~\ref{txt:phot_comp}), which caused larger uncertainties
in the determination of their R$_{\rm eff}$ as well.

The parameter $\mu_{\rm eff}(B)$ is more difficult to compare
(middle panel of Figure~\ref{fig:struct_comp}), since only for
pure exponential disks our defined value (the mean inside R$_{\rm eff}$) and
the ISIB96 parameter ($\mu$ of the narrow annulus at R$_{\rm eff}$) can be
converted into each other through a simple relation.
Even when accounting for this ``disk'' difference, the scatter
is still significant and shows some systematic deviations on both high and
low surface brightness levels.
The typical uncertainty of the photometric calibration of the
photographic plates used in
the work by ISIB96 is $\sim$0\fm25 \citep{Sprayberry96}. This
implies that the uncertainty of their  $\mu_{\rm eff}(B)$ is at least
at this level. Besides the term related to
$\sigma_{\mu_{\rm eff}}$ should be added. Thus, $\sigma_{\mu_{\rm eff}}$ is
calculated with the equation
\begin{equation}
\sigma_{\mu_{\rm eff}} = \sqrt{ \sigma_{\rm tot}^2 + \Big( 2.17 \cdot \frac{\sigma_{R_{\rm eff}}}{\rm R_{\rm eff}} \Big)^2},
\end{equation}
where $\sigma_{\rm tot}$ is the uncertainty in the total/observed magnitude
and $\sigma_{R_{\rm eff}}$ is the uncertainty for $R_{\rm eff}$.
As discussed above, the error in R$_{\rm eff}$ for our data was estimated
directly. For ISIB96 data it was estimated for a purely exponential disk
as 0.15 of R$_{\rm eff}$.

The combined error bars for the differences in $\mu_{\rm eff}$
are shown in the middle panel of Figure ~\ref{fig:struct_comp}. The great
majority of the data from our SDSS parameters and those from ISIB96 are
consistent within 2$\sigma$ of the combined error ($\sim$0\fm8--1\fm0).
However, for about 20\% of the galaxies the differences are larger.
As follows from Section~\ref{txt:phot_comp}, the uncertainties of our
photometry are significantly lower than those from ISIB96.
Thus the total uncertainties for the differences in $\mu_{\rm eff}$
will be defined mainly by uncertainties from \citet{Sprayberry96}.
Therefore, large deviations
(i.e., more than twice the combined $\sigma$, or $\sim$0\fm8--1\fm0)
should be attributed both to the mentioned photometry problems for some
objects in ISIB96 and to differences in the definition of $\mu_{\rm eff}$.

\begin{figure}
    \begin{center}
    \epsscale{1.0}
    \includegraphics[angle=0,width=18cm]{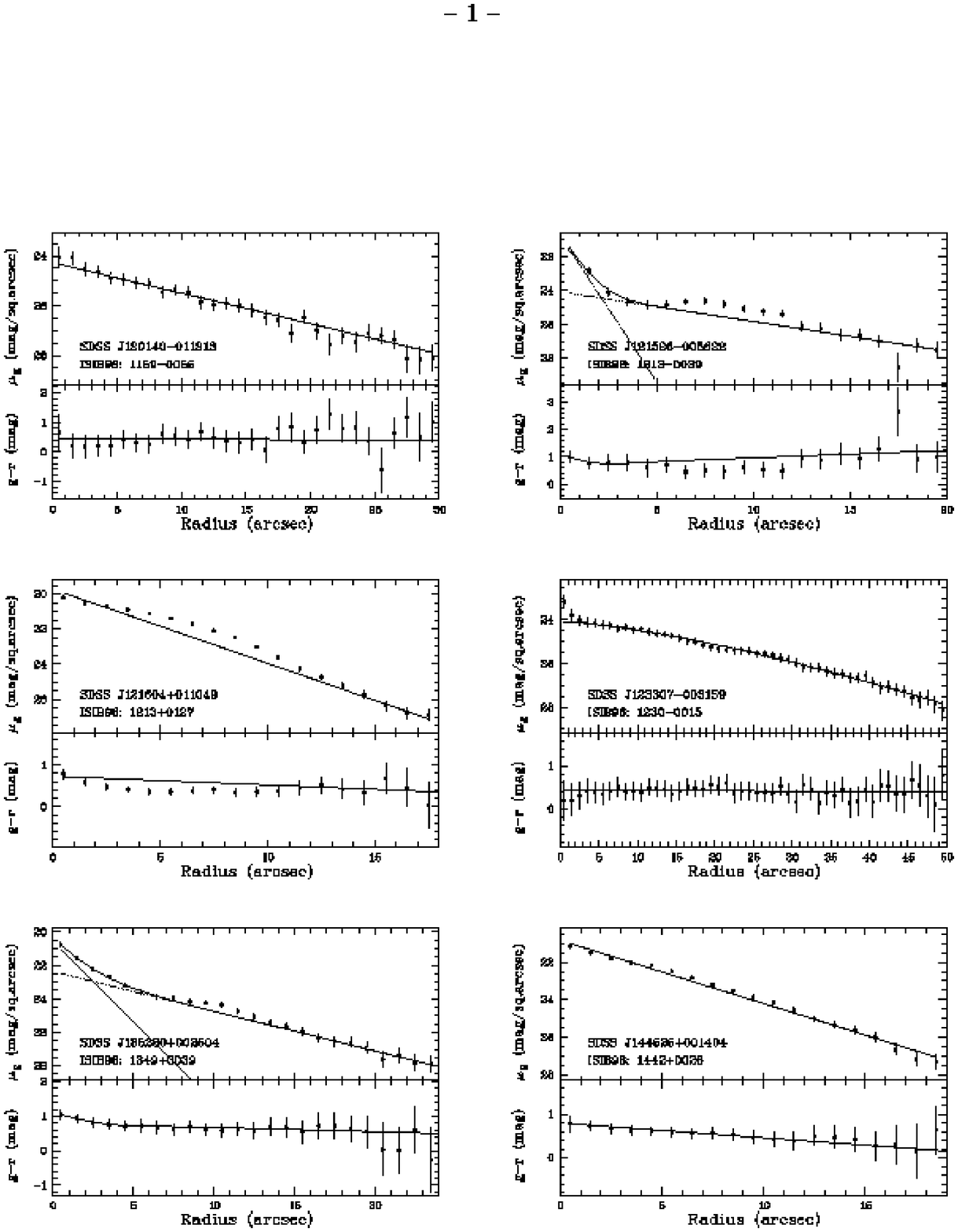}
    \caption{
The SDSS $g$-band SBPs and $(g^*-r^*)$ color radial profiles for
six of the galaxies from the test sample.
Three kinds of profiles are shown: pure exponential
(1159$-$0055, 1213$+$0127, 1442$+$0026),
disk and bulge (1212$-$0039 and 1349$+$0039), and
S\'ersic profiles (1230$-$0015).
The different lines in the SBPs panels show the model fit distributions to
the data, including bulge (if any), disk, and the total
light.
The lines in the $(g^*-r^*)$ color panels show the model $(g^*-r^*)$ color
distributions derived as the difference between the model
$g^*$ and $r^*$ distributions.
    \label{fig:SBPs}}
    \end{center}
\end{figure}

\begin{figure}
    \begin{center}
    \epsscale{1.0}
    \includegraphics[angle=0,width=18cm]{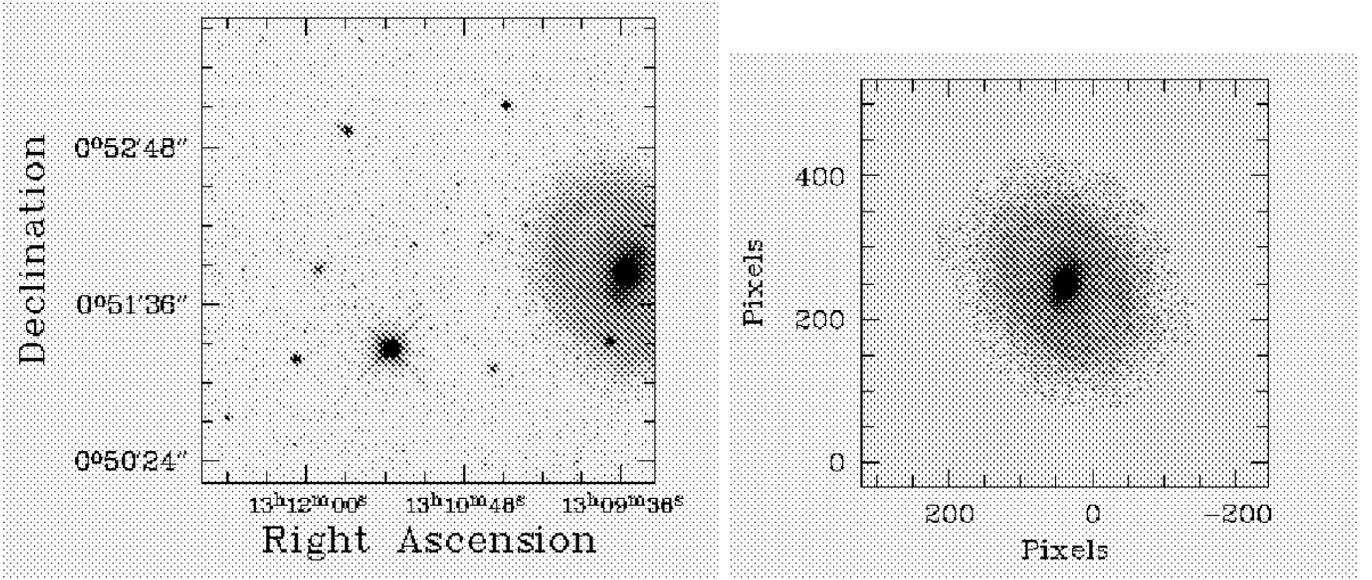}
    \caption{
Galaxy NGC~4996 (J130931$+$005125), detected as an additional, 
non-ISIB96 object
in one of the fields of the test sample. 
The galaxy (as seen in the SDSS field) is shown in the
left panel.  It is truncated by the field border.  In order to ``restore''
the truncated part, we assumed point symmetry.  The resulting ``restored''
image is shown in the right panel.  The difference  between
the magnitude calculated within the original mask and the one calculated
for the ``restored'' image is $\Delta m(g^*) =$~0\fm22.
    \label{fig:examples_bilt}}
    \end{center}
\end{figure}

\begin{figure}
    \begin{center}
    \epsscale{1.0}
    \includegraphics[angle=-90,width=15cm]{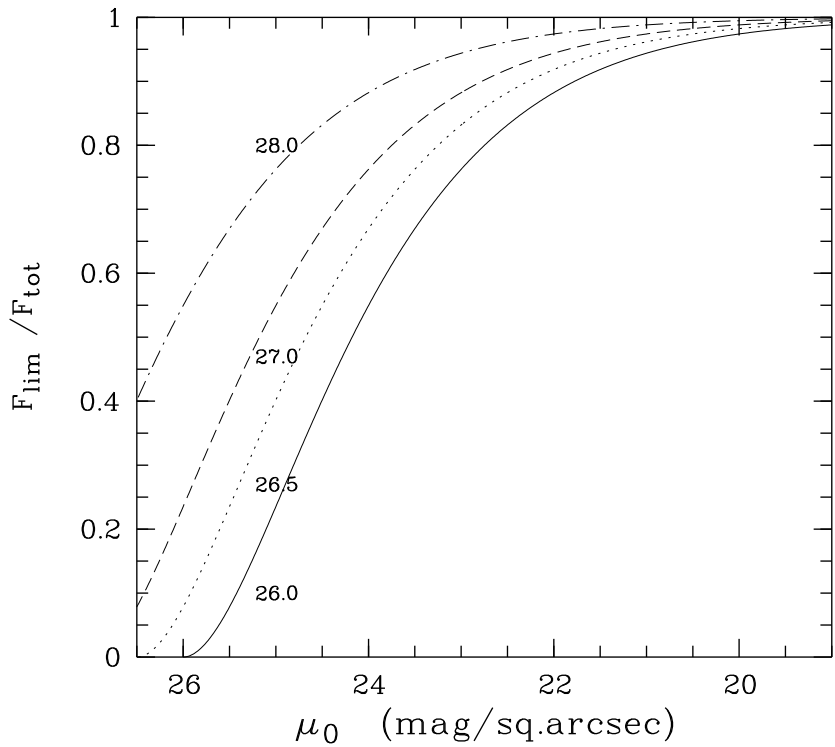}
    \caption{
Filter-independent ratio  $F_{\rm obs}/F_{\rm tot}$ of the total
galaxy flux to that observed within the limiting surface brightness
$\mu_{\rm lim}$ versus the
central surface brightness $\mu_{\rm 0}$.
    \label{fig:ratio_model}}
    \end{center}
\end{figure}

\begin{figure}
    \begin{center}
    \epsscale{1.0}
    \includegraphics[angle=0,width=11cm,clip=,bb=160 140 450 735]{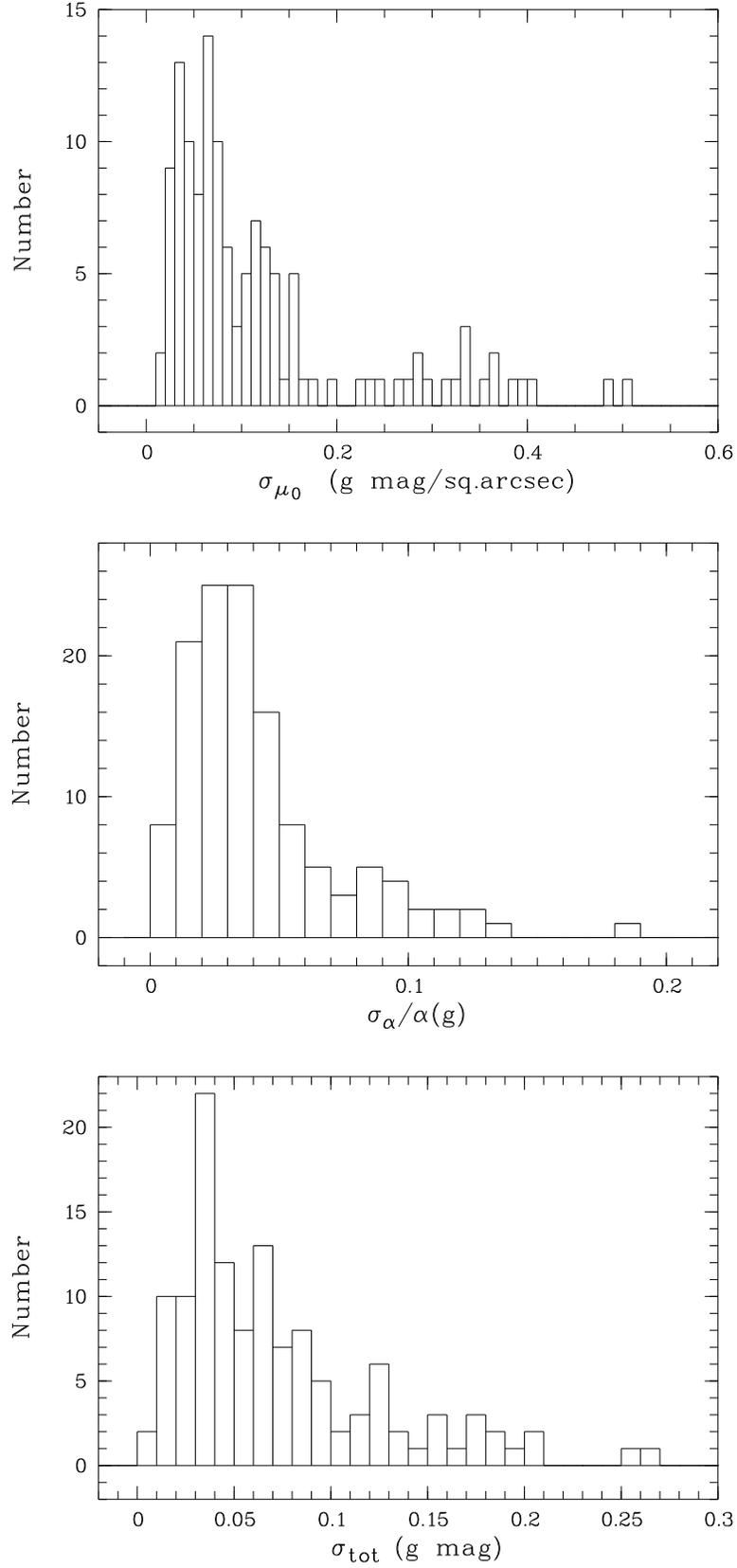}
    \caption{
    The distributions of the errors  of the exponential
    fitting parameters $\mu_{\rm 0}$ and $\alpha$ for the galaxies of our
    sample in the $g$-band    (top and middle panels,
    respectively) and the distribution of the uncertainties in their $g$-band 
    total magnitude (bottom panel), derived with the procedure outlined
    in Section~\ref{txt:total_mag}.
    \label{fig:model_errors}}
    \end{center}
\end{figure}

\begin{figure}
    \begin{center}
    \epsscale{1.0}
    \includegraphics[angle=-90,width=18cm]{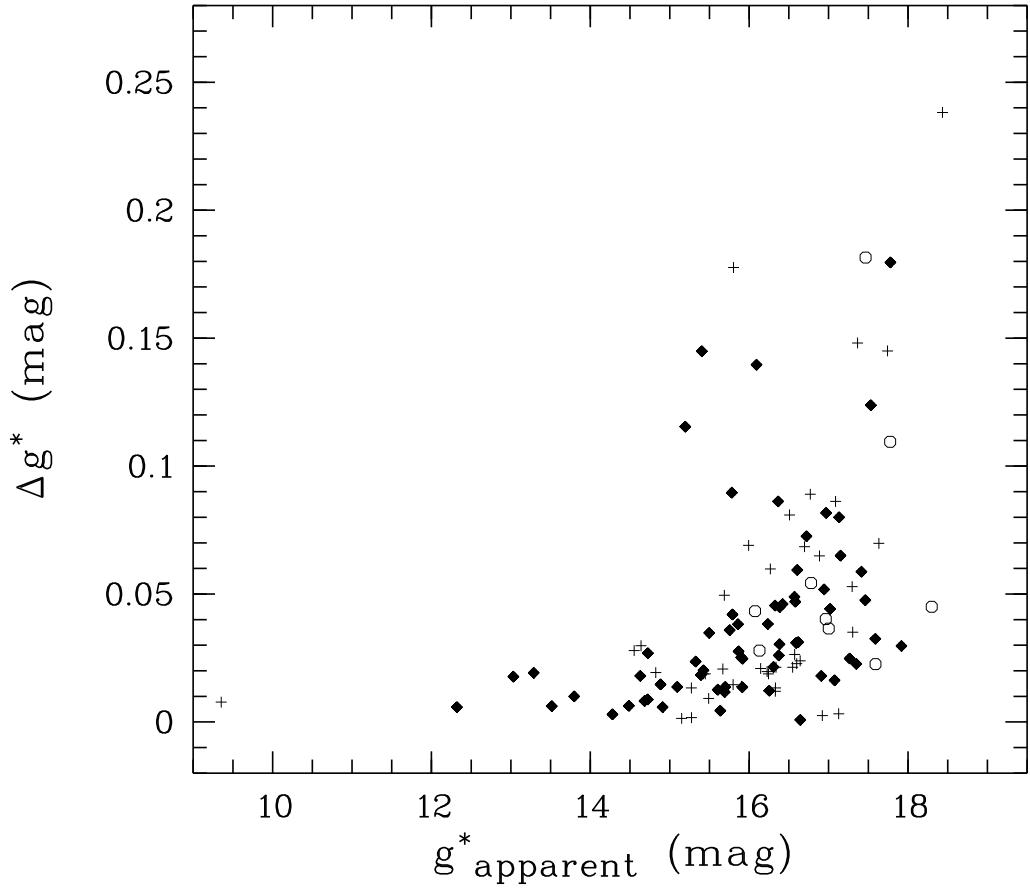}
    \caption{
The difference between the apparent and the total $g$-band
magnitudes versus the apparent magnitude for our
sample galaxies. The different
symbols are the same as in Figure~\ref{fig:ratio_model}
and denote the type of SBP fitting.
    \label{fig:appar_total}}
    \end{center}
\end{figure}

\begin{figure}
    \begin{center}
    \epsscale{1.0}
    \includegraphics[angle=0,width=18cm,clip=,bb=75 275 525 600]{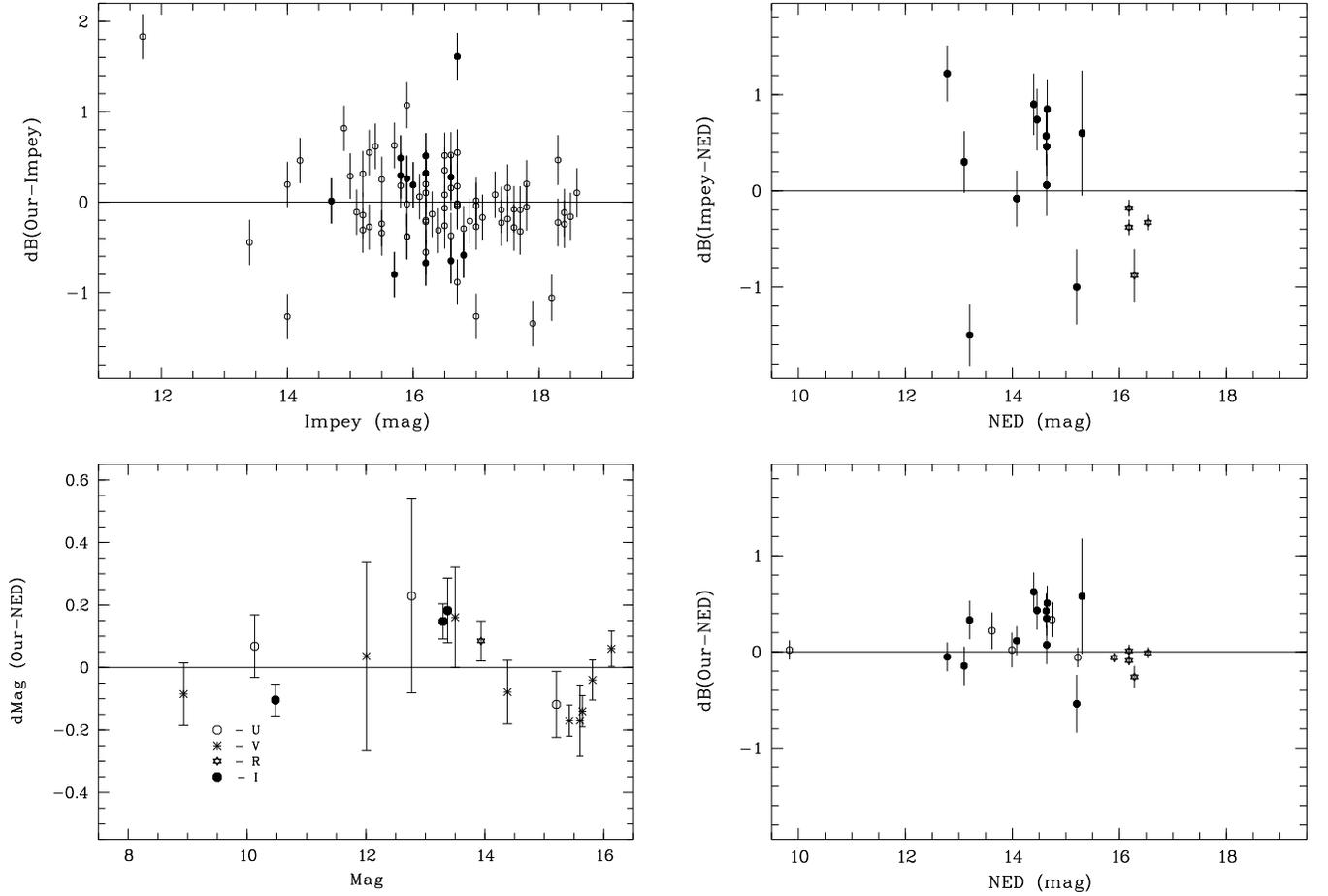}
    \caption{
Comparison of our photometry
(transformed to $UBVR_cI_c$ magnitudes using the
equations from \cite{SDSS_phot1})
with $B_{\rm tot}$ from ISIB96 and with magnitudes for several
bright galaxies retrieved from NED.
{\it Top-left panel:} The difference $\Delta B$(Our$-$Impey) versus
$B_{\rm tot}$ from ISIB96. The filled circles denote galaxies
that, according to ISIB96, had CCD photometry as opposed to have been
measured on photographic plates.
{\it Top-right panel:} $\Delta B$(Impey$-$NED) versus $B$(NED). The filled
circles show all galaxies in ISIB96 and NED in common.
The comparison of ISIB96 photometry with the magnitudes from
\citet{Salzer89} is shown with asterisks.
{\it Bottom-left panel:} Comparison of our photometry (transformed to
$UVR_cI_c$) with the magnitudes from NED.
{\it Bottom-right panel:} $\Delta B$(Our$-$NED) versus $B$(NED).
The galaxies from ISIB96 are shown by filled circles, while the
additional galaxies that we found in the test sample fields
are shown by open circles.
The comparison of our photometry with the magnitudes from \citet{Salzer89}
is shown with asterisks.
    \label{fig:phot_comp}}
    \end{center}
\end{figure}

\begin{figure}
    \begin{center}
    \epsscale{1.0}
    \includegraphics[angle=0,width=11cm,clip=,bb=160 140 450 735]{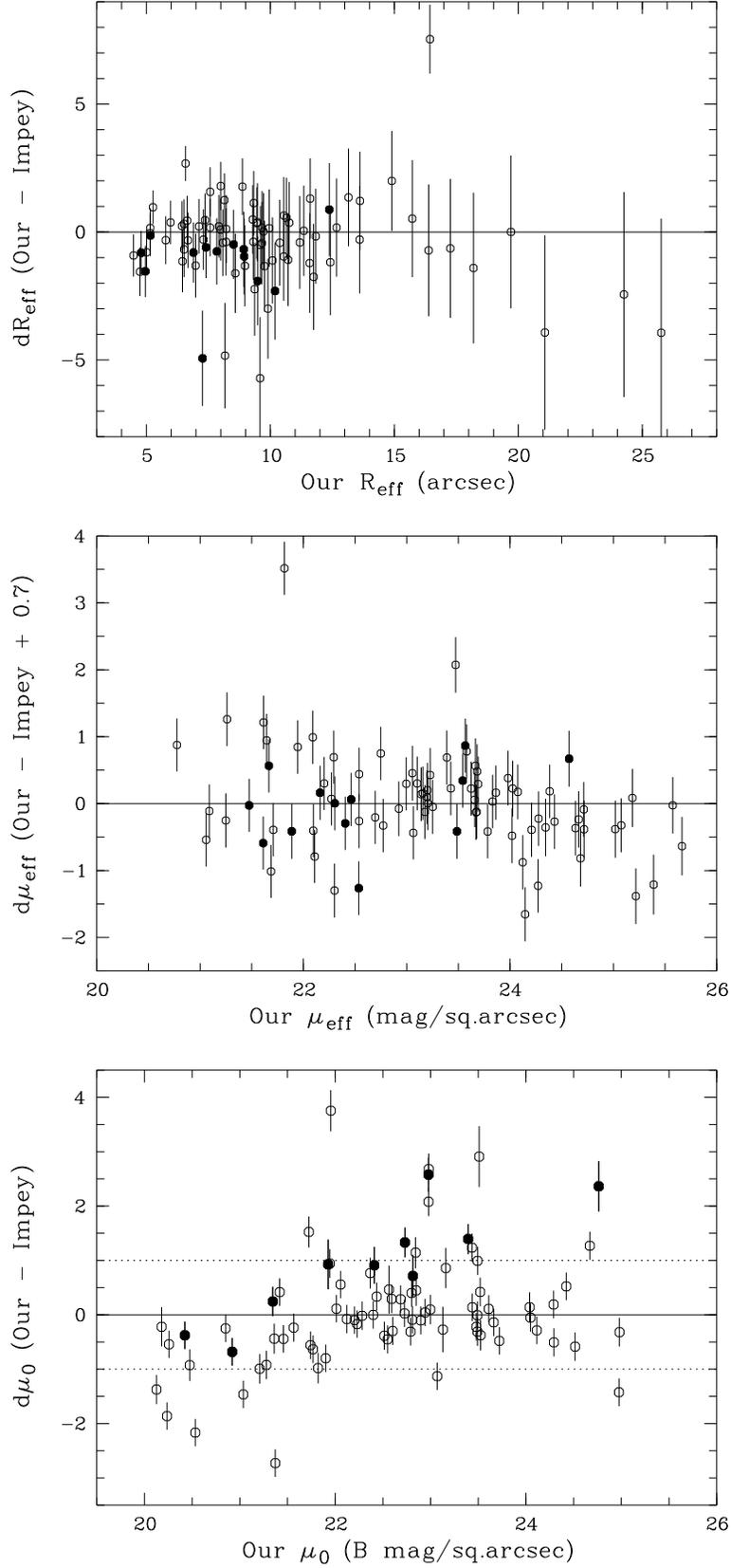}
    \caption{
Comparison of our structural parameters R$_\mathrm{eff}(g^*)$,
$\mu_{\rm eff}$, and $\mu_{\rm 0}$ with those from ISIB96
for the test sample galaxies. Our original $g$ and $r$-band 
$\mu_{\rm eff}$ and $\mu_{\rm 0}$ were transformed to $B$-band magnitudes.
The error bars corresponding to
the total r.m.s. uncertainties of shown differences.
Filled circles denote galaxies that according to ISIB96
had parameters derived from CCD photometry (as opposed to photographic
plates).
    \label{fig:struct_comp}}
    \end{center}
\end{figure}

Of greater concern, however, are the rather large differences
apparent in the comparison of the model-dependent central surface
brightness $\mu_{\rm 0}(B)$ that is shown in the bottom panel
of Figure~\ref{fig:struct_comp}.
For each point the total uncertainty includes the 0\fm25 from
ISIB96 and the uncertainty that was defined in course of our fitting procedure.
In reality, the average uncertainty for $\mu_{\rm 0}(B)$ given by ISIB96
should probably be
at least two times larger than average uncertainty of the total magnitude.
We show, therefore, the average combined 2$\sigma$ level as 1\fm0.
However, even with this 2$\sigma$ uncertainty level about 23\% of objects
exhibit larger deviations.
In particular, in many cases, for galaxies with a central
luminous component such as a bulge or bar, the 
parameter $\mu_{\rm 0}(B)$ from ISIB96
was significantly brighter than ours.
Since our surface photometry is very deep, in most of these cases we could 
see that the value presented by ISIB96 refers to the $\mu_{\rm 0}(B)$
of this central brighter component, but not to the underlying disk,  in spite
of ISIB96's description of their procedure for the
derivation of $\mu_{\rm 0}(B)$.
In Figure~\ref{fig:SBPs} we show, as an illustration, SB profiles of two
such cases.  According to ISIB96, the galaxy 1212$-$0039 has
$\mu_{\rm 0}(B)$=22.40 mag~arcsec$^{-2}$, while 1349$+$0039 
has $\mu_{\rm 0}(B)$=20.30 mag~arcsec$^{-2}$.
For seven galaxies we found that ISIB96 give a $\mu_{\rm 0}(B)$
significantly lower ($>$1\fm0) than what we derive. We show SBPs of
two of these galaxies in Figure~\ref{fig:SBPs} to illustrate that
according to our data it would be very difficult to get such a large error.
According to ISIB96,  1213$+$0127 and 1442$+$0026 have 
$\mu_{\rm 0}(B)$=22.10 mag~arcsec$^{-2}$ and 24.10 mag~arcsec$^{-2}$,
respectively.

Only for one galaxy, 1226$+$0105 from the test sample, we found
a SBP published in the literature.
According to ISIB96, its
$\mu_{\rm 0}(B)$=20.90 mag~arcsec$^{-2}$, which differs considerably
from our $\mu_{\rm 0}(B)$=22.94$\pm$0.11 mag~arcsec$^{-2}$.
On the other hand, according to \citet{Sprayberry95}, its
uncorrected $\mu_{\rm 0}(B)$=23.00$\pm$0.14 mag~arcsec$^{-2}$, which is very
close to our value.

To summarize, we exploited the superb quality of the SDSS imaging data
to re-measure the structural and photometric properties.  We then
compared our results with the published data for the test sample galaxies.
For a substantial number of galaxies the estimates of some
parameters that are important for further analysis are improved in
comparison to the values derived in the original work.

\section{Discussion}
\label{txt:disc}

\subsection{The selection thresholds of our method}
\label{txt:limpar}

In a later paper we will address issues concerning
the resulting selection function,
and completeness via Monte-Carlo simulations of the selection process.
However, some useful estimates for the selection thresholds of our method
can be derived using very simple
models and through the measured parameters of the detected galaxies.

\subsubsection{Limiting isophote}

The use of highly homogeneous SDSS images obtained in drift-scan mode
and of a very simple primary selection criterion (total area) allow us to
easily estimate the limiting isophote parameter of our selection procedure.
However, the use of combined $gri$ and filtered images to search for
galaxies makes the estimation of this criterion more complicated.
As a first step, we tried to estimate the limiting isophote in the $g$ filter
($\mu_{\rm lim}(g^*)$) using the background level and noise estimation
on $g$ band images before combining, after combining, and after
filtering of the combined frames.
This is possible only because our background level for images in the $g$
filter and in the combined images is the same as follows 
from our description of
the steps for aligning  and combining the $gri$ images 
(see Section~\ref{txt:method_alig} for details).
We carried out such estimates for several randomly selected
fields.  They show that we have a mean limiting surface brightness
3$\sigma$ isophote $\mu_{\rm lim}(g^*) \approx$~26.4 mag~arcsec$^{-2}$.
Unfortunately, devising a similar procedure 
of estimating $\mu_{\rm lim}$ for other bands is less straightforward.

Next, we tried to estimate the
limiting surface brightness isophote analytically using the following
equation for a
pure exponential surface brightness profile \citep{AllenShu79}:

\begin{equation}
\label{equ:allen_shu}
m_{\rm tot} = \mu_0 + 5 \cdot \log_{10} \Bigl[\frac{0.7349}{\theta_{\rm lim}}(\mu_{\rm lim} - \mu_0)\Bigr]
\end{equation}
In this equation $\theta_{\rm lim}$ is the limiting angular diameter
(in arcsec) of the galaxy with an exponential SBP
at the limiting surface brightness isophote $\mu_{\rm lim}$.
Since $\mu_0$ and $\mu_{\rm lim}$ in this equation refer to the same band
as $m_{\rm tot}$, we can apply equation (\ref{equ:allen_shu}) to any
photometric band of interest.
The expression for $\theta_{\rm lim}$ may be used for any
angular-limited sample. For example, \citet{Impey88} applied it
to analyze selection effects based on the structural parameters
for a sample of LSB galaxies found in the Virgo Cluster.

In our case $m_{\rm tot}$ for a purely exponential SBP
can be calculated using the fit parameters
and equation (\ref{equ:disk_tot}).
The limiting surface brightness is defined as the level above which we
detected our objects.  It can therefore be determined
by comparison with a grid of model curves from equation (\ref{equ:allen_shu})
with $\mu_{\rm lim}$ as a parameter.
Since we defined our limiting angular size $\theta_{\rm lim}$
precisely as 24\arcsec\ (Section~\ref{txt:method_detect}), 
one needs to vary only the limiting surface brightness $\mu_{\rm lim}$
of these models in order to put the model curves
on the plot of $\mu_{0}(g^*)$ versus total $g^*$ magnitude
for underlying disks.
The respective lines for the $g$ band are shown in
Figure~\ref{fig:model_limits},
where we plotted them for three limiting surface brightnesses
$\mu_{\rm lim}$ = 25.0, 26.0, and 26.5 mag~arcsec$^{-2}$.
As evident from the figure, both the test sample galaxies
(filled diamonds) and additionally selected galaxies (crosses) define
the unique model curve (as expected from the selection procedure),
corresponding to the limiting surface brightness $\mu_{\rm lim}(g^*) =$~26.5
mag~arcsec$^{-2}$. There are no galaxies to the left of this curve. All are
situated either to the right of it or fall (within the uncertainties of their
parameters) right on top of this curve.

This estimated limiting surface brightness isophote for $g$ is
even slightly fainter than our previous rough estimate
$\mu_{\rm lim}(g^*) =$~26.4 mag~arcsec$^{-2}$, but the
difference between these two values is not large.
Since the second estimate is a global one
(i.e., it allows us to produce an estimate for all data
simultaneously), we will adopt this value as the estimate of
$\mu_{\rm lim}(g^*)$ hereafter.
A similar analysis was performed for the $r$ and $i$ bands as well.
The resulting threshold values are
$\mu_{\rm lim}(r^*) =$~26.2 mag~arcsec$^{-2}$ for the $r$ band
and $\mu_{\rm lim}(i) =$~25.9 mag~arcsec$^{-2}$ for the $i$ band.
To compare our limiting parameters with those from previous studies
we transformed $\mu_{\rm lim}(g^*)$ and $\mu_{\rm lim}(r^*)$ to
$\mu_{\rm lim}(B)$ with the equations from \cite{SDSS_phot1}.
Thus, our limiting surface brightnesses correspond to
$\mu_{\rm lim}(B) \sim$~26.9 mag~arcsec$^{-2}$ in the standard Johnson B-band.
This value is slightly deeper in comparison to the claimed value of ISIB96
$\mu_{\rm lim}(B) \sim 26$ mag~arcsec$^{-2}$.

\subsubsection{Scale lengths}

\begin{figure}
    \begin{center}
    \epsscale{1.0}
    \includegraphics[angle=-90,width=15cm]{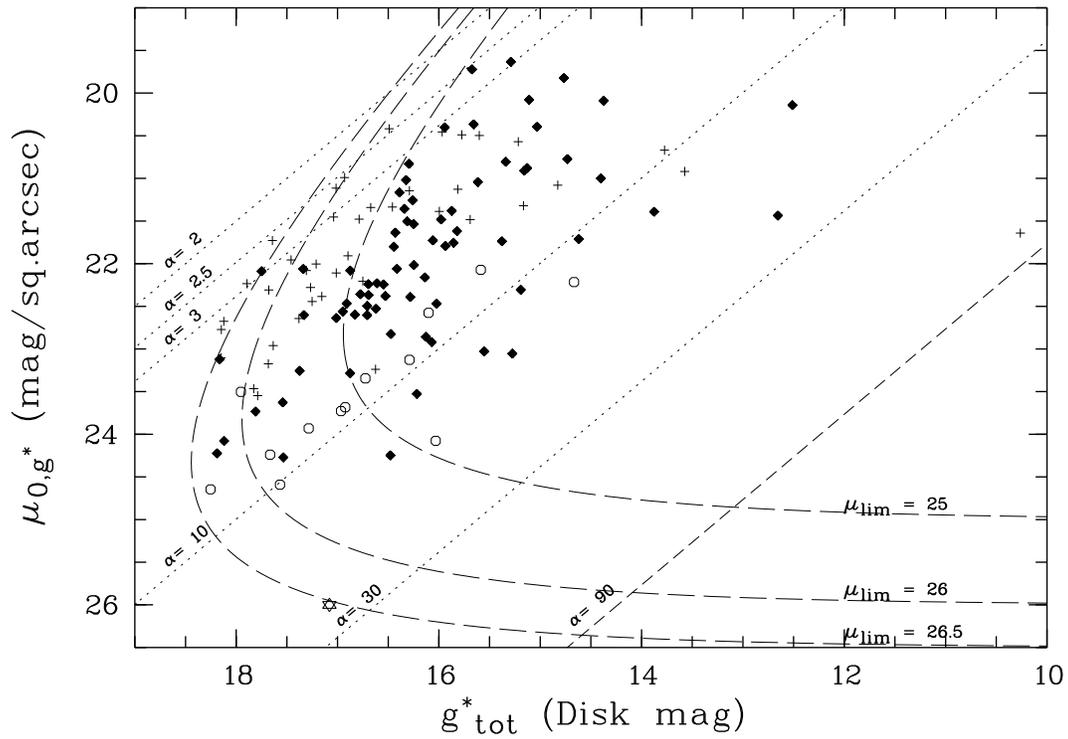}
    \caption{
The total disk $g^*$-magnitude versus the central surface
brightness $\mu_0$ for all detected galaxies.
The long-dashed curves are the envelopes of the selection function
defined by the limiting isophotes $\mu_{\rm lim}$ labelled in the legend
and the limiting
angular size at those isophotes, $\theta_{\rm lim}$ = 24\arcsec.
They were calculated using equation~\ref{equ:allen_shu}.
Isolines (dotted) with constant scalelength $\alpha$ are overplotted as well.
Filled diamonds denote the galaxies from the ISIB96 test sample.
Open circles mark the galaxies from this test sample
with S\'ersic SBPs ($n > 1$).
Crosses show the additional galaxies  found in the 
SDSS fields with the same selection criteria as used to retrieve the 
test sample.
The star symbol at $g^*_{\rm tot} \approx$17\fm0 shows the location of the
faintest ghost image detected with our software.
    \label{fig:model_limits}}
    \end{center}
\end{figure}

Based on the plot in Figure~\ref{fig:model_limits} we can also
discuss the distribution of scale lengths $\alpha$ of the detected galaxies.
The dotted straight lines drawn in this plot show the positions of exponential
disks from equation (\ref{equ:disk_tot})
with scale lengths of $\alpha$ = 2\arcsec, 2\farcs5, 3\arcsec,
10\arcsec, 30\arcsec, and 90\arcsec.
As expected from the limiting isophotes in Figure~\ref{fig:model_limits},
the most compact galaxies that our software should be able to detect are
objects with $\alpha$ = 2\arcsec, if their $\mu_0(g^*)$ is brighter than
or equal to 20 mag~arcsec$^{-2}$.
However, although our sample contains galaxies with bright centers,
the five most compact galaxies actually detected (both from the
test sample as well as new detections) have $\alpha (g^*) \approx$ 2\farcs5.
Thus, from this rather limited sample we can infer that our programs
can confidently detect disk galaxies with scale lengths
$\alpha_{\rm lim}(g^*) \gtrsim 2\farcs5$.
As one can see, most of the sample galaxies have $\alpha (g^*)$ ranging from
3\arcsec\ to 10\arcsec. Only 11 galaxies ($\sim$8\%) have $\alpha (g^*) > $
10\arcsec. Two galaxies in the sample are found to have $\alpha (g^*) > $
20\arcsec. One of them, with $\alpha (g^*)$ = 23\arcsec, is a galaxy from the
test sample. So, the real fraction of extended objects in our sample is
seemingly limited not by our methods, but by their presence in the test
sample.
The evidence for the effective detection of extended galaxies comes from
the largest and the brightest galaxy in our sample -- NGC~3521 with
$\alpha (g^*) \approx$~75\arcsec, which was absent in the test sample.

Finally, we conclude that our software is capable to detect
galaxies with scale lengths $\alpha (g^*)$ at least up to $\sim$1\arcmin\ on
SDSS images.
Of course, the limited size of the SDSS fields (12\arcmin$\times$10\arcmin)
begins to play a significant role in the selection procedure for such
extended galaxies. However, even for such extended galaxies our
programs will detect at least ``the tip
of iceberg'' with sufficiently high probability.
It is also worth noting that on the SDSS frame containing NGC~3521,
whose final mask had the size of $\sim$1/2 of the whole frame,
our programs found two more galaxies, one of which (1103$+$0007) exists also
in the test sample. As for galaxies with $\alpha \geq$ 90\arcsec\
(dashed line in Figure~\ref{fig:model_limits})
one has to work with the much larger 
SDSS mosaic frames (see Stoughton et al.\ 2002 for details)
to ensure their detection.

\subsubsection{Central surface brightnesses}

The expected limiting $\mu_{\rm 0}$ also can be estimated from
the plot in Figure~\ref{fig:model_limits}. It of course depends on
the galaxy scale length. In particular, for galaxies with
$\alpha (g^*) \geq$~25\arcsec\ our software should detect galaxies as dim as
$\mu_0(g^*) \geq$~26.0 mag~arcsec$^{-2}$ (which corresponds to
$\mu_0(B) \geq$~26.3 mag~arcsec$^{-2}$). Since these estimates account only
for pure exponential disks, any additional light
from bulge, bar, or H\,{\sc ii} regions will improve the possibility to
detect such a faint disk.

\begin{figure}
    \begin{center}
    \epsscale{1.0}
    \includegraphics[angle=-90,width=18cm]{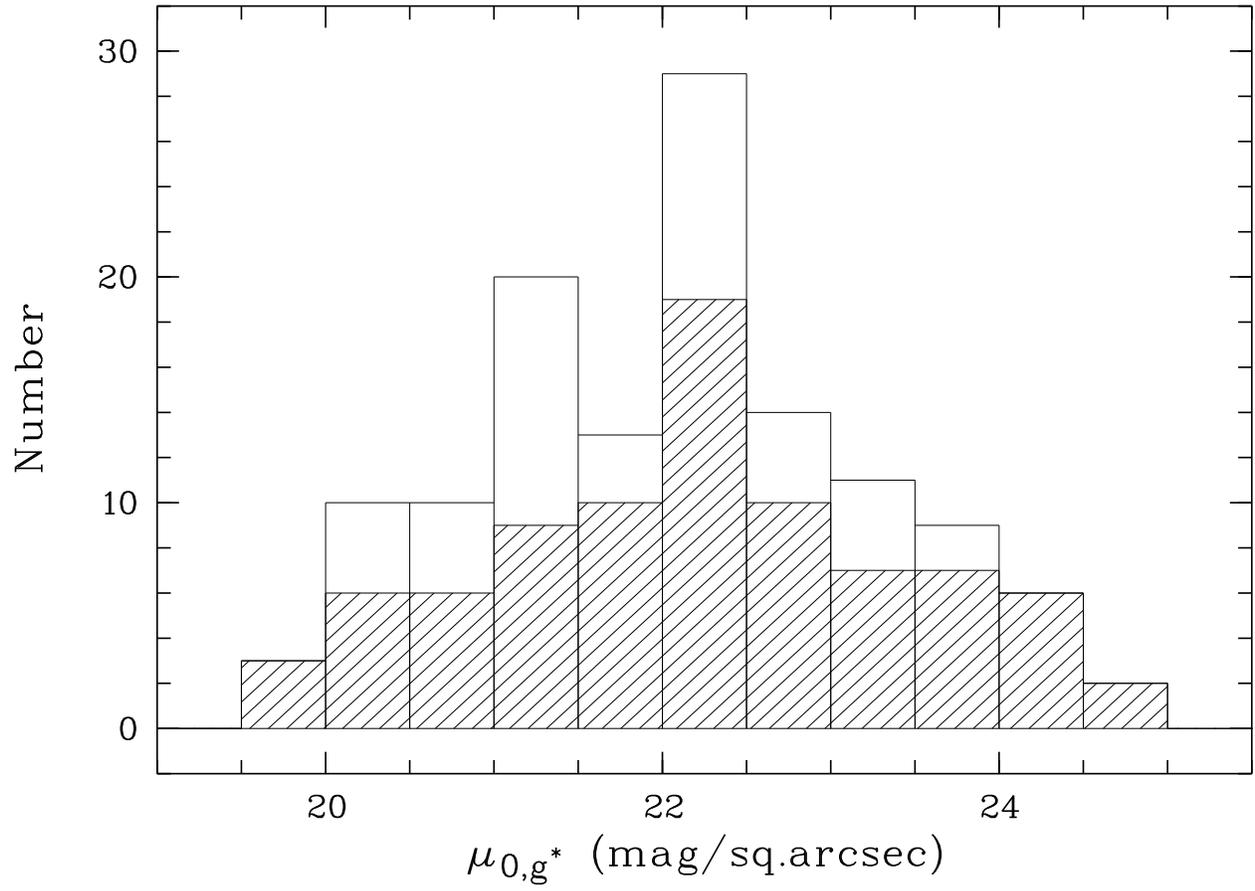}
    \caption{
The histogram of the observed central surface brightnesses $\mu_0(g^*)$.
The galaxies from ISIB96 are shown as hashed bins.
    \label{fig:observed_mu0}}
    \end{center}
\end{figure}

\begin{figure}
    \begin{center}
    \epsscale{1.0}
    \includegraphics[angle=-90,width=15cm]{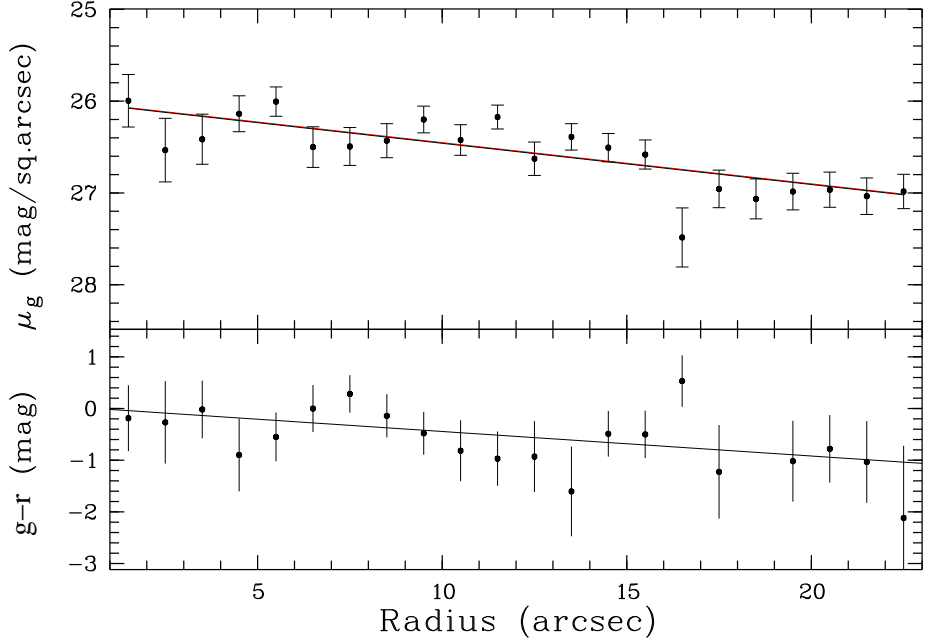}
    \caption{
 The SBP in the $g$-band, and the $(g^*-r^*)$ radial profile of the faintest
 ghost
 image detected by our programs in  one of the test sample fields.
    \label{fig:ghost_g}}
    \end{center}
\end{figure}

The distribution of $\mu_0(g^*)$ for all studied galaxies is shown
in Figure~\ref{fig:observed_mu0}.
All objects, both from the test sample as well as the additional
detections, 
fall in the range of $19.5 <\mu_0(g^*) < 25.1$ mag~arcsec$^{-2}$.
Based only on the properties of these galaxies one can hardly 
judge the possibility to detect fainter disks with our software.
However, for further analysis we can use the fact that several objects as dim
as $\mu_0(g^*) =$~26.0 mag~arcsec$^{-2}$ were detected during the preliminary
process of extended object identification.
In the subsequent checks they were recognized as artifacts
(ghosts of bright stars). Their parameters, however, indicate that
if these objects were real galaxies, galaxies this faint will very likely
be detected by our software.
In Figure~\ref{fig:ghost_g} we show for
illustration the surface brightness profile of one of the ghost artifacts
with $\mu_0(g^*) =$~26.00 mag~arcsec$^{-2}$  and $\alpha(g^*) =$~24\farcs18.
These ``disk'' parameters correspond to a total magnitude of
$g^*_{\rm tot}=$~17\fm08.
A real galaxy with such parameters will appear just on the border
of the allowed region on diagram $\mu_0(g^*)$ versus $g^*_{\rm tot}$
in Figure~\ref{fig:model_limits} (star symbol).

\subsubsection{Edge-on galaxies}

It is worth noting that our selection criteria produce a bias against
edge-on galaxies. Since we separate objects with the total area within
the limiting isophote greater than some minimal threshold, the lower limit for
their major-axis diameter of 24\arcsec\ is valid only for round, face-on
galaxies. For elongated galaxies, the lower limit for this parameter
varies as
\begin{equation}
D_{\rm lim} = \frac{24\arcsec}{\sqrt{b/a}}
\end{equation}
Thus, for the most elongated
galaxies ($b/a \sim$0.1) we will detect all objects with major-axis
diameter larger than 76\arcsec.

\subsubsection{Objects with S\'ersic profiles}

Since in order to describe SBPs of galaxies with general S\'ersic
profiles one needs one more parameter, the discussion of galaxy selection
effects becomes more complicated
when indices of $n >$~1 (equation \ref{equ:exp_profile}) have to be used.
We detected all galaxies of this type present in the test sample when
applying our homogeneous selection criteria  based on the size on
the  limiting isophote. They are shown in Figure~\ref{fig:model_limits}
by open circles.
However, for the same values of $\alpha$ and
$\mu_0$ such objects have total magnitudes fainter than
objects with $n =$~1. The amount of the total flux decrease
depends on value of $n$. For example,
$F_{\rm n}/F_{\rm n=1} =$~0.8 for $n =$~1.67 and
$F_{\rm n}/F_{\rm n=1} =$~0.58 for $n =$~5.
Thus, $n >$~1 shifts effectively such objects to the left in
Figure~\ref{fig:model_limits}. This implies, e.g., that some of these
galaxies with SBP parameters falling close to the limiting model
curve with $\mu_{\rm lim}(g^*) =$~26.55 mag~arcsec$^{-2}$ will remain
undetected.
This will cause some additional problems for the analysis of completeness
in the future survey.  This issue may be properly addressed via
modelling of the S\'ersic profile disk detection.

\subsection{First scientific results}

Our test sample from ISIB96 was chosen to allow us to optimize our search
criteria such that we are capable of retrieving known LSB galaxies.  For
the development of our programs and algorithms we confined ourselves to the
subset of SDSS fields containing
the galaxies from the test
sample.  These fields are non-contiguous and are drawn from a much larger
area.  Owing to our biased area selection we cannot draw conclusions 
regarding the statistical distribution and density of LSB galaxies, nor 
can we easily assess the overall completeness.  However, the detection of
numerous additional galaxies that fall in the same parameter space as the
ISIB96 sample indicates that the SDSS imaging data have the potential to 
significantly increase the number of known LSB galaxies (see also Section
\ref{txt:future}).  In any case, even our current, limited survey has lead
to a number of interesting scientific findings, some of which will be 
discussed briefly below. 

\subsubsection{Galaxy SDSS~J140831$-$000737: two disks with distinct
spiral structure?}
\label{txt:J140831}

The depth and homogeneity of the SDSS imaging data allow us to study
processes such as large-scale star formation in the faint outer regions 
of galactic disks \citep{Ferg98}, which have been little studied so far.
For example, a galactocentric radius $R>R_{25}$ (where $R_{25}$
is defined by the $B = 25$ mag isophote) corresponds to an
isophotal level $\mu(g^*) \sim 24.6$ mag~arcsec$^{-2}$ for $(g^*-r^*) = 
0\fm5$, and an even brighter level for redder $(g^*-r^*)$ colors.  These
levels are much brighter than our limiting isophotal level of 
$\mu(g^*) = 26.55$ mag~arcsec$^{-2}$.
The study of star formation processes in the faint outskirts of 
disk galaxies provides
a unique insight into the nature of the star formation law, preconditions
for star formation, and star formation thresholds in regions 
that are characterized by physical environments (such as
lower metallicities and longer dynamical time scales, etc.) that are very
different from those typical of the bright, inner regions of disks
\citep[see, e.g.,][]{Nog01}.

\begin{figure}
    \begin{center}
    \epsscale{1.0}
    \includegraphics[angle=0,width=18cm]{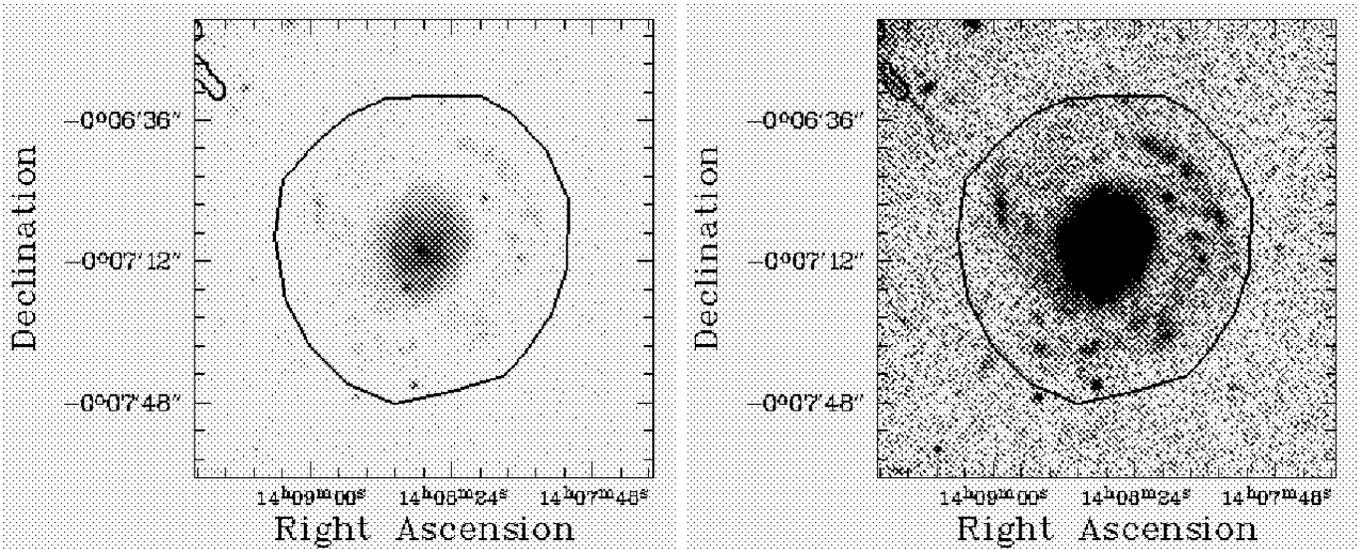}
    \caption{
     Two presentations of galaxy SDSS~J140831$-$000737
     to highlight its unusual structure.
     {\it Left panel:} A combined SDSS $g, r$, $i$ image with
     the isophotal selection level superimposed.
     This image is plotted with relative low and high intensity levels
     of $-1\sigma$ and 100$\sigma$, respectively.
     Spiral arms in the central, ``inner'' disk 
     are clearly recognizable on the combined image,
     as are bulge and bar-like structures.
     {\it Right panel:} The same image is plotted, but with 
     relative low and high intensity values of $-1\sigma$ and 5$\sigma$.
     Patchy outer spiral arms can be clearly seen at this level of
     contrast.  Previous studies missed the ``outer disk'' visible in the
     SDSS data and detected only the inner portion of this galaxy.
    \label{fig:outer_disk}}
    \end{center}
\end{figure}

An interesting galaxy offering such an opportunity is SDSS~J140831$-$00073
(1405$+$0006 in ISIB96).  This galaxy
has an unusual structure reminiscent of the multi-component disks mentioned by
\citet{BF93}.
The SBP of the disk of this galaxy is approximated best by two exponentials,
corresponding to an ``inner''  and an ``outer'' disk.
This galaxy was classified by ISIB96 as Sb(r).  On the SDSS images, a bulge,
a bar, and different spiral patterns in the ``inner'' disk and in the 
``outer'' disk are clearly seen (Fig.\ \ref{fig:outer_disk}).
The ``inner'' disk is characterized by
$\mu_0(B) = 21.1$ mag~arcsec$^{-2}$ and $\alpha = 1.6$ kpc.
At a surface brightness of $\mu(g^*) \sim 24.5$ mag~arcsec$^{-2}$
($R >$~15\arcsec), the ``inner'' spiral arms disappear.  This surface
brightness limit also corresponds to the apparent size
of this galaxy given in previous studies, and to the outer limit of the
``inner'' disk.
On the SDSS images we found a new subsystem of spiral arms fainter than 
$\mu(g^*) \sim 25.0$ mag~arcsec$^{-2}$.  The contrast-enhanced image
(plotted with $-1 \sigma$ to $5 \sigma$, where $\sigma$ is the noise of
the sky background) is shown in the right panel of
Figure~\ref{fig:outer_disk}.  The limiting isophotal level above which this
galaxy was detected  corresponds to $\mu(g^*) \sim 26.55$ mag~arcsec$^{-2}$
(plotted as a solid line in Figure~\ref{fig:outer_disk}), yielding a
scale length $\alpha$ for the ``outer'' disk of $\sim 7$ kpc.
The central surface brightness for the ``outer'' disk is
$\mu_0(B) = 24.5$ mag~arcsec$^{-2}$.
The flux ratio of the ``inner'' and the ``outer'' disk, 
m$_{tot}^{ext.disk}$/m$_{tot}^{int.disk}$, is approximately 1.2.

While the detailed nature of this galaxy remains uncertain at this point,
we may be seeing an example of the class of galaxies with ``stepped disks''
described in \citet{MG97}.  It is interesting that each
of the two disk components exhibits its own distinct spiral pattern.  
Planned long-slit spectroscopy will tell us more about the rotational
properties of SDSS~J140831$-$00073 and the metallicities of the ``inner''
and ``outer'' disks. 
\citep{ImpBurk01} show that the H\,{\sc i} profile of this system
is very narrow 
(W$_{50} = 92$ km s$^{-1}$ and W$_{20} = 112$ km s$^{-1}$).
\citet{MG97} discuss the possibility that this kind
of galaxy may have a very low effective viscosity, allowing outlying
gas to retain its initial angular momentum.
A possibly similar case is the galaxy NGC~6946 described in \citet{Ferg98},
which also shows spiral arms in its outskirts that do not appear to be 
connected with its ``inner'' arms.

\subsubsection{Morphological types and concentration index}

Our photometric programs calculate a number of global
morphological parameters for every galaxy.  Some of these may be
useful for morphological galaxy classifications
(see earlier discussions by, e.g., \cite{Kent85, KWO86, FR99, 
Shimasaku2001, Strateva2001}).
A particularly useful parameter is the concentration index ($C$ hereafter),
defined as the
ratio of the radii containing 90\% and 50\% of a galaxy's light.
For the classical de Vaucouleurs profile, $C$ is $\sim$5.5,
and for pure exponential disks, $C \sim 2.3$.  These values are valid 
for the idealized seeing-free case, which we are approximating due to
the limiting angular size $\theta_{\rm lim} \gg PSF$ that we impose
in this work.

\begin{figure}
    \begin{center}
    \epsscale{1.0}
    \includegraphics[angle=-90,width=18cm]{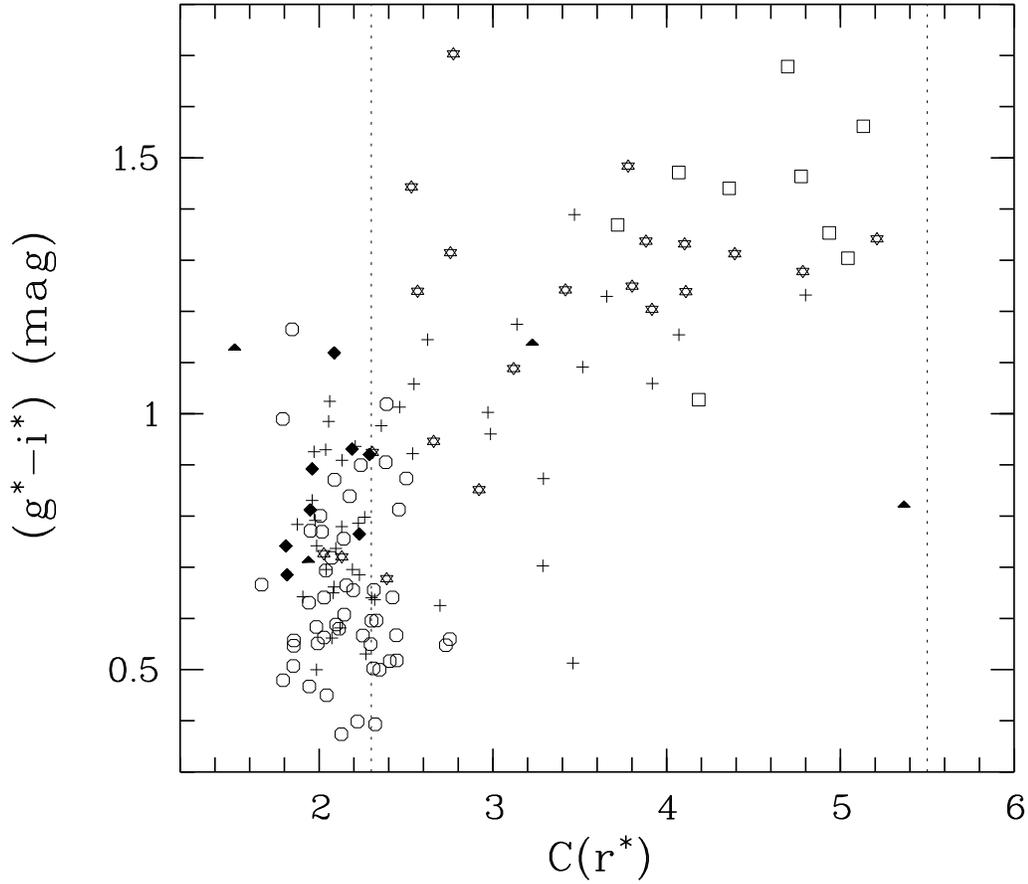}
    \caption{
The relation between the color index $(g^*-i^*)$ and the concentration
index in the $r^*$-band, C($r^*$).
Open squares denote E galaxies.
Asterisks denote early-type galaxies (S0 -- Sa).
Crosses indicate galaxies of ``intermediate'' type (Sb -- Sc).
Open circles stand for late-type galaxies (Sd -- Irr and dI).
Filled lozenges denote dEs.
Filled triangles denote interacting galaxies.
The values of the concentration index for classical de Vaucouleurs profiles
(5.5) and for pure exponential disks (2.3) are shown by dashed lines.
    \label{fig:conc_index}}
    \end{center}
\end{figure}

In Figure~\ref{fig:conc_index} the relation
between the color index $(g^*-i^*)$ and the concentration index in the 
$r$-band, C($r^*$), is plotted.
We selected the color index $(g^*-i^*)$ for two reasons: (1) as a 
robust index from the point of view of signal to noise, and (2) since
it provides a good temperature baseline considering the wavelength
separation between the two bands.
ISIB96 and NED list morphological classifications for the galaxies in
our sample, which we revised in a number of cases based on the 
SDSS image material.
The adopted classification is shown in column 10 of Table~\ref{tbl-1}
(see Section~\ref{txt:test}).
Based on this classification we divided all galaxies in our sample
into following categories (barred types included) shown in
Figure~\ref{fig:conc_index}:
(a) early-type spiral galaxies (S0 -- Sa);
(b) intermediate-type spirals (Sb -- Sc);
(c) late-type galaxies: Sd, Sm, and various Irr and dI;
(d) dEs; and
(e) interacting galaxies.
We added E galaxies (open squares) in this Figure to see overall
distribution of all different morphological types.
The theoretical 
values for the concentration index for classical de Vaucouleurs profiles
and for pure exponential disks are shown by dashed lines there too.

The bimodality in the distribution of galaxies
noticed by \citet{Strateva2001} is clearly visible in 
Figure \ref{fig:conc_index}:
Early-type galaxies are located in the parameter region
$(g^*-i^*) > 0\fm9$ and $C(r^*) >2.6$, and
late-type galaxies are characterized by 
$(g^*-i^*) \le 1\fm2$ and $C(r^*) < 2.6$.
Intermediate-type spirals are found in both branches of the distribution.
Our empirical separation at $C(r^*) = 2.6$ for these ``branches'' is in excellent
agreement with the one found by \citet{Strateva2001}, $C(r^*) = 2.63$.
But using $C(r^*)$ as the only morphological separation criterion
has certain drawbacks that lead to ambiguities:
Additional substructure such as spiral arms and bulges in spiral
galaxies or H\,{\sc ii} regions in irregular galaxies will change both 
$C$ and color and often act in opposite ways.  
We plan to use the large sample of galaxies that will result from the
application of our programs to larger areas of the SDSS 
(Section~\ref{txt:future}) to devise an improved,
more detailed classification scheme.

A few more comments on Figure~\ref{fig:conc_index}:
(1) Our empirical upper limit for the concentration index is $\sim 5.2$,
close to the theoretical value for the classical de Vaucouleurs profile 
($C = 5.5$).  Our upper limit differs from the value $\sim 3.7$ calculated 
based on Petrosian radii by \citep{Shimasaku2001, Strateva2001}
as given in the SDSS photometric database. 
This difference is caused by the method employed for the calculation of
Petrosian magnitudes by the standard SDSS software (e.g., \cite{Strauss02}),
which systematically underestimates Petrosian magnitudes
not only for galaxies with pure de Vaucouleurs profiles, but for all galaxies
with bulges.  The amplitude of this shift depends on the bulge-to-disk ratio 
\citep{Strauss02}).
(2) Interacting galaxies as complex systems may be located in any part
of this diagram and cannot easily be recognized.
(3) Disk systems are characterized by $C(r^*) < 2.3$.  Shifts to
lower values of $C$ occur when additional
light from spiral arms and/or H\,{\sc II} regions contribute. 
(4) Practically all dE systems are located below $C(r^*) = 2.3$ as well.
This is in agreement with the theoretical expectations since 
most of these galaxies have
SBPs with $n > 1$ (equation (\ref{equ:exp_profile})).

\begin{figure}
    \begin{center}
    \epsscale{1.0}
    \includegraphics[angle=0,width=18cm]{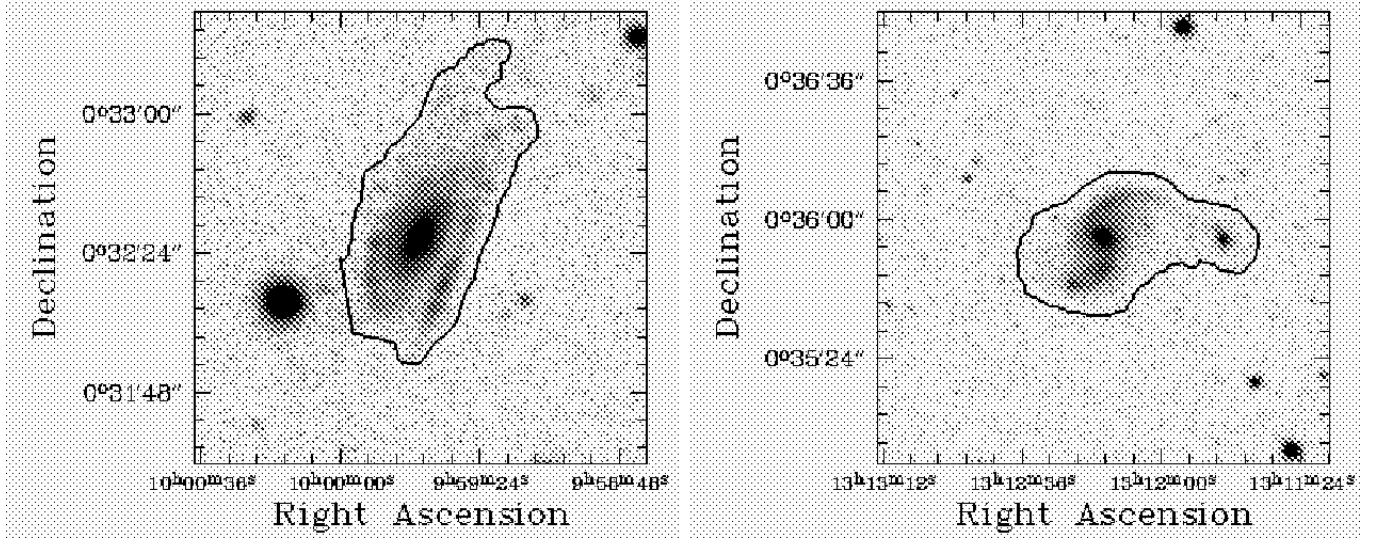}
    \caption{
     Combined $g, r$, and $i$ images of the galaxies
     SDSS~J101142+003520 and SDSS~J131215+003554 with
     superimposed isophotes that correspond to the threshold level of
     the selection criterion.
     Both galaxies together with SDSS~J140831-000737
     (see Figure~\ref{fig:outer_disk}) were found as unusual objects
     using the $(g^*-r^*)$ vs.\ $C$ plot in the region of the parameters
     $(g^*-i^*) < 0\fm9$ and $C >2.8$ (see Figure~\ref{fig:conc_index}).
     Galaxy SDSS~J101142+003520 ({\it left panel}) looks lopsided,
     and galaxy SDSS~J131215+003554 ({\it right panel})
     exhibits either a bright H\,{\sc ii} region, an interacting galaxy,
     or a superimposed background galaxy at the far end of its western
     spiral arm.
     The small object was observed spectroscopically by the SDSS.  It 
     has a velocity of 14406 km s$^{-1}$ and a strong H\,{\sc ii}
     emission line spectrum.
     A spectrum for the galaxy SDSS~J131215+003554
     has not yet been obtained, preventing further interpretation.
    \label{fig:compl_morph}}
    \end{center}
\end{figure}

The $(g^*-i^*)$ versus $C(r^*)$ contains five outliers at
$(g^*-i^*) < 0\fm9$ and $C(r^*) >2.8$. One of them is the
interacting system SDSS~J133031$-$003613 (1327$-$0020).
The other four galaxies have unusual morphologies:
The ``stepped disk'' galaxy SDSS~J140831$-$000737 
($(g^*-i^*) = 0\fm51$ and $C(r^*) = 3.46$) was described in the
previous Section~\ref{txt:J140831}.  The
galaxy SDSS~J101142$+$003520 ($(g^*-i^*) = 0\fm85$ and $C(r^*) = 2.92$)
looks lopsided and is shown in Figure~\ref{fig:compl_morph} (left
panel).
The galaxy SDSS~J131215$+$003554 ($(g^*-i^*) = 0\fm70$ and $C(r^*) = 3.29$)
may be interacting with what appears to be 
a smaller galaxy seen at the extension of the
tip of its western spiral arm (Figure~\ref{fig:compl_morph}, right panel).  
Alternatively, this seemingly smaller
object may be a background galaxy.
The galaxy SDSS~J095939$+$003227 ($(g^*-i^*) = 0\fm87$ and $C(r^*) = 3.29$)
may have an outer blue disk of low surface brightness.
All of these ``unusual'' galaxies identified in the $(g^*-i^*)$, $C(r^*)$
diagram are characterized by unusual processes in their faint outer regions
and are interesting targets for future follow-up studies.

\subsubsection{Giant LSB galaxies}
\label{txt:1405+0006}

An interesting aspect of this pilot study is the identification of 10
luminous distant galaxies with considerable bulge
and/or bar components.
The underlying disks are low-surface-brightness disks with
disk scale lengths in the range of $\sim$7.5 to 13 kpc.
These galaxies fall into the category of the so-called
giant LSB galaxies, or `cousins' of Malin~1 \citep{Bothun87}.
\citet{Sprayberry95} define these objects as LSB galaxies that meet a
``diffuseness index''  criterion: $\mu_{\rm 0,c} + 5\cdot\log(\alpha) > $27.6
(adopting a Hubble constant of H$_{0} =$ 75~\kms~Mpc$^{-1}$). Here
$\mu_{\rm 0,c}$ is the extrapolated, deprojected, and corrected for
cosmological dimming $B$-band disk central brightness.
The properties of giant LSB galaxies
are summarized in the paper by \citet{Sprayberry95}.
Currently, only 16 galaxies of this type are known.  Their H\,{\sc i}
properties were recently studied by \citet{Matthews2001}.

\begin{figure}
    \begin{center}
    \epsscale{1.0}
    \includegraphics[angle=-90,width=18cm]{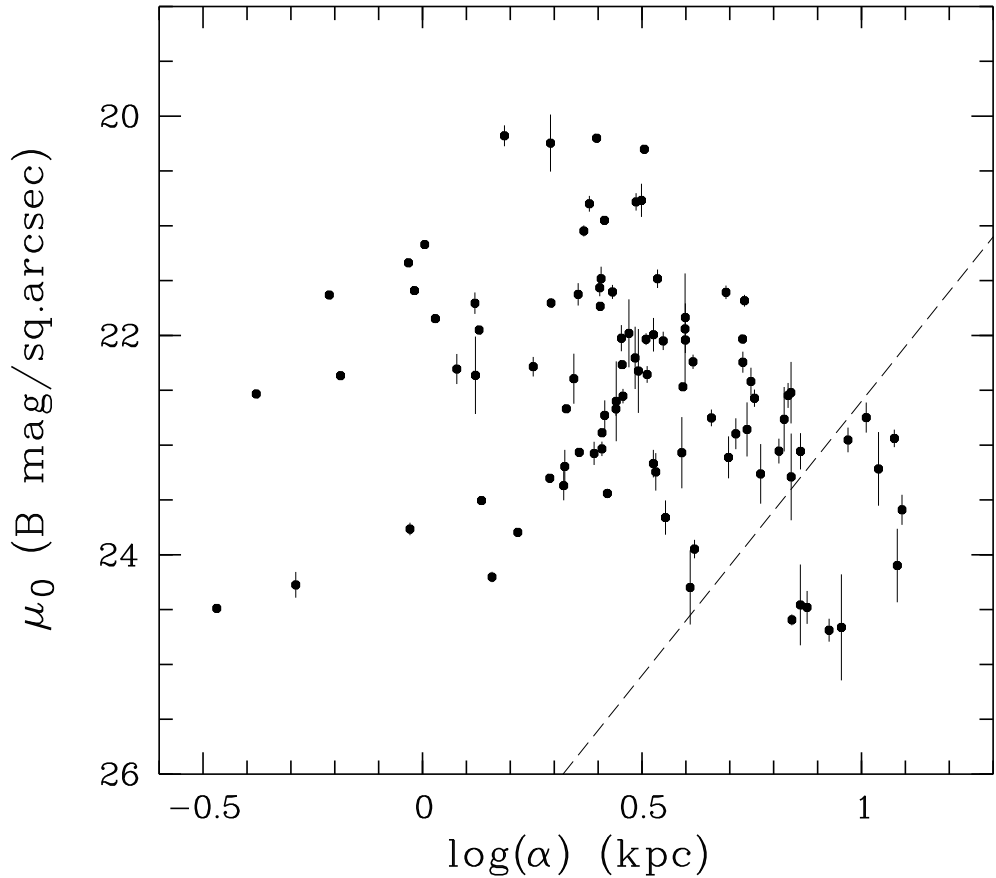}
    \caption{
Central surface brightness $\mu_0$ as a function of scale length $\alpha$
for our sample.
The dashed line marks the cutoff in ``diffuseness index'' of
$\mu_{\rm 0,c}(B) + 5\cdot\log(\alpha) = 27.6$, which is
used to define a giant LSB disk \citep{Sprayberry95}.
    \label{fig:giant_LSBs}}
    \end{center}
\end{figure}

We present  a list of eleven giant LSB galaxy
candidates (Table \ref{tbl:giant_LSBs}),
which includes only one previously known galaxy
of this type -- 1226+0105, described by \citet{Sprayberry95}.
The location of these galaxies in a plot of scale length versus
central surface brightness is shown
in Figure~\ref{fig:giant_LSBs}. 
As \citet{Bothun1997} emphasize, giant LSB galaxies are quite enigmatic from
the point of view that they normally formed their spheroidal component,
but no conspicuous stellar disk ever formed around their bulge. Improved
statistics for these objects will lead to a 
better understanding of the relationship between their fundamental parameters
and the conditions/processes that led to their formation. The
high detection rate of giant LSB galaxy candidates with our programs
promises substantial progress for future systematic studies.

\subsection{A large LSB survey with SDSS data}
\label{txt:future}

The main goal of our project is to create a big and uniform LSB
sample on the base of SDSS imaging data.
We plan to extend our analysis to the
entire sky region from which the test sample described here was extracted,
i.e., the EDR runs 752 and 756 (7272 SDSS fields) comprising 
$\approx$228 deg$^2$ \citep{Stoughton2002}.  This area is 76 times larger
than the total area of the sky region
analyzed in the current study.
This future sample will cover a sufficiently large region of the
sky to provide us with better statistics and to decrease the
effects caused by large-scale
fluctuations of galaxy density. 

We plan to use our final sample to conduct a thorough study of 
the properties of LSB galaxies, including the
analysis of their luminosity function.  We will also investigate the
space density of galaxies as a function
of central surface brightness, for which we will need to create a 
volume-limited sample.
This implies that we need redshifts for all our selected galaxies
from the whole region of the two EDR runs.
In this sense, the selection threshold on total magnitude of
about $r$ = 18\fm0, which we derive from the
estimations performed using Figure~\ref{fig:model_limits},
looks very promising. It is rather close to the SDSS galaxy spectroscopy
threshold magnitude of $r$ = 17\fm77.
Accounting for the 2dF redshift survey information already available
in the discussed region \citep{Colless01}, we expect that only for a small
fraction of the detected LSB galaxies we will need 
additional redshift determinations.

\section{Summary and conclusions}
\label{txt:summ}

In this paper we pursued the goal to develop an effective method to
search for LSB and dwarf galaxies in the SDSS image database based on
a test sample from the galaxy catalog of ISIB96.
We developed our own programs to search for galaxies with large angular
sizes from SDSS images. The reliability of these programs was
carefully tested using a training subsample of galaxies from the catalog
of ISIB96. This allowed us to choose our selection criteria such
that we achieved very high efficiency in recovering galaxies from the
test sample, limited only by
cases when a candidate galaxy falls too close to a very bright star or galaxy.
The dimmest LSB galaxy from the test sample detected so far has
$\mu_{0}(B) \sim$ 25~arcsec$^{-2}$.
Based on the tests discussed in the previous sections, we estimate our
detection limits to be  $\mu_{0}(g^*) =$ 26\fm0 -- 26\fm5 arcsec$^{-2}$,
which will need to be verified in our planned studies.
Furthermore, our new method, which applies predefined objective
selection criteria, resulted in the detection of 42 (47\%)
additional galaxies with similar parameters within the test fields.

A continuation of this work is planned within a region with a
76 times larger area. On the one hand, this will allow us to refine
the determination of the detection limit of our method,
and on the other hand is expected to create a new sample of $\sim$10$^3$
galaxies.

Based on the work described in our current paper, the following 
conclusions can be drawn:

\begin{enumerate}
       \item
	Dealing directly  with the SDSS catalog database parameters, we
	found it difficult to reliably identify LSB galaxies.
	To overcome these difficulties 
        we developed our own method, which we plan
	to apply to the entire SDSS imaging data.
	A program package intended to search for galaxies with large
	angular size was created and carefully tested, and is
        described in the present paper in detail.
	These programs allowed us to find 87 of a training set of 
        90 galaxies from the Impey et al.\ (1996) catalog, distributed over
	93 SDSS fields.
	That is, the efficiency of the method is $\sim$96.5\%.
	Our new software also resulted in the discovery of
        42 additional galaxies
	in the same SDSS fields, whose parameters are similar to
	those of the test sample objects.
       \item
	We also created programs that produce independent
	photometry of all selected galaxies within the studied SDSS images.
        The comparison of our photometry for galaxies found twice in
        overlapping fields of the same SDSS run shows good repeatability.
	All differences are within 1--3$\sigma$ of the photometric 
        uncertainties, indicating that our programs are very robust.
	We compared our photometry with published data in various
	bands for several galaxies in common. This shows the
	external consistency of our magnitudes to be within $\sim$0\fm1
	for different bands.
	We demonstrate that our programs allow us to obtain accurate 
        photometry for galaxies as bright as $B$=10\fm0 from SDSS images.
       \item
        The results of this study show that the SDSS images
        can be efficiently used to search for LSB galaxies
        down to at least $\mu_{0}(g^*) \sim$ 25.0~mag~arcsec$^{-2}$.
	Our software has been shown to be sensitive to galaxies with 
        scale lengths $\alpha$
	in the range between 2\farcs5 and 75\arcsec and
	with central surface brightness $\mu_0(g^*)$
	in the range between 19.5 and 26.0 mag~arcsec$^{-2}$; it may work
        fine over an even larger range.
       \item
	The results obtained for the test sample also demonstrate the 
        potential of the SDSS imaging data for the study of unusual
        LSB galaxies:
	\begin{itemize}
	\item We found a galaxy (SDSS~J140831$-$000737) 
	    with a ``stepped disk'' structure that looks like two
	    exponential disks, an ``inner'' and an ``outer'' one,
	    akin to galaxies described in \citet{MG97}.
            Both disks show their own distinct spiral pattern.
	\item We showed that galaxies with unusual faint outer
	    structure may be found on the base of a
	    color $(g^*-r^*)$ vs. concentration index $C(r^*)$
            diagram.  We confirm the usefulness of this diagram for
            crude morphological galaxy classifications.
	\item We found ten new giant LSB galaxy candidates in
	      addition to the 15 previously known giant LSB galaxies
	      or Malin~1-like objects, underlining the potential of 
              the SDSS for the study of these enigmatic LSB objects.
	\end{itemize}
\end{enumerate}

In subsequent papers, we will explore the selection function of our
galaxy sample, luminosity functions, and the detailed properties of our
galaxies and of their environment.  Also, we will continue our survey
for LSB galaxies using SDSS data and targeted follow-up observations.
Our studies will help to improve the census of volume-selected LSB
galaxies and of their evolutionary status.

\acknowledgments

S.A.P. and A.G.P. acknowledge financial support and hospitality of MPIA 
during part of this work. The authors are grateful to the referee
for the useful questions and suggestions which helped to improve the paper
presentation.

Funding for the creation and distribution of the SDSS Archive has been 
provided by the Alfred P. Sloan Foundation, the Participating Institutions, 
the National Aeronautics and Space Administration, the National Science 
Foundation, the U.S. Department of Energy, the Japanese Monbukagakusho, 
and the Max Planck Society. The SDSS Web site is http://www.sdss.org/. 

The SDSS is managed by the Astrophysical Research Consortium (ARC) for the
Participating Institutions. The Participating Institutions are The
University of Chicago, Fermilab, the Institute for Advanced Study, the 
Japan Participation Group, The Johns Hopkins University, Los Alamos 
National Laboratory, the Max-Planck-Institute for Astronomy (MPIA), 
the Max-Planck-Institute for Astrophysics (MPA), New Mexico State 
University, University of Pittsburgh, Princeton University, the United 
States Naval Observatory, and the University of Washington.

This work is based in part on observations in the framework of the ``Calar
            Alto Key Project for SDSS Follow-up Observations''
	    \citep{Gre01} obtained at the German-Spanish
            Astronomical Centre, Calar Alto Observatory,
            operated by the Max Planck Institute for Astronomy,
            Heidelberg, jointly with the Spanish National
            Commission for Astronomy.

This research has made use of the
NASA/IPAC Extragalactic Database (NED) which is operated by the Jet
Propulsion Laboratory, California Institute of Technology, under contract
with the National Aeronautics and Space Administration.

\clearpage

\setcounter{qub}{0}
\begin{deluxetable}{rlllrllllrll}
\rotate
\tabletypesize{\tiny \bf}
\tablecaption{The general parameters of large-angular-size galaxies identified
in the fields of the test subsample. \label{tbl-1}}
\tablewidth{0pt}
\tablehead{
\colhead{\#} &\colhead{SDSS name} &\colhead{Impey name} & \colhead{$\alpha(2000.0)$} & \colhead{$\delta(2000.0)$} &
\colhead{R,C,F} & \colhead{$B$} & \colhead{$B-V$} & \colhead{M$_B$} &\colhead{V$_{\rm hel}$} & \colhead{Type} & \colhead{Other names} \\
& & & & & & \colhead{mag} & \colhead{mag} & \colhead{mag} & \colhead{km/s} & & \\
&\colhead{(1)} & \colhead{(2)} & \colhead{(3)} & \colhead{(4)} &  \colhead{(5)} &
\colhead{(6)}  & \colhead{(7)} & \colhead{(8)} & \colhead{(9)} & \colhead{(10)} & \colhead{(11)}
}
\startdata
\qq&SDSS J095146$+$001950 &            &09:51:46.80 &$+$00:19:50.09 & 756   4   214&  17.13$\pm$0.05&  0.99& $-$20.51& 25291& Sc(r)  &                         \\
\qq&SDSS J095138$+$002211 &0949$+$0036 &09:51:38.88 &$+$00:22:11.27 & 756   4   214&  17.31$\pm$0.06&  0.58& $-$14.70&  1897& Im     &                         \\
\qq&SDSS J095938$+$003047 &            &09:59:38.16 &$+$00:30:47.29 & 752   5    42&  16.72$\pm$0.04&  0.67& $-$20.35& 19466& Sa     &                         \\
\qq&SDSS J095939$+$003227 &            &09:59:39.60 &$+$00:32:27.57 & 752   5    42&  16.11$\pm$0.03&  0.72& $-$20.15& 13386& SBc    &UGC5370                  \\
\qq&SDSS J095939$+$003511 &0957$+$0049 &09:59:39.60 &$+$00:35:11.89 & 752   5    42&  15.95$\pm$0.03&  0.92& $-$21.15& 19762& Irr?   &                         \\
\qq&SDSS J095959$+$003816 &0957$+$0052 &09:59:59.52 &$+$00:38:16.02 & 756   5   228&  16.88$\pm$0.05&  0.55& $-$18.69&  9735& Sm     &                         \\
\qq&SDSS J095930$-$003745 &            &09:59:30.24 &$-$00:37:45.46 & 752   2    42&  17.45$\pm$0.06&  0.80& $ $     &      & Sa     &                         \\
\qq&SDSS J100439$+$001653 &            &10:04:39.36 &$+$00:16:53.38 & 756   4   236&  17.07$\pm$0.05&  0.85& $-$17.60&  6464& E?     &                         \\
\enddata
\end{deluxetable}

\setcounter{qub}{0}
\begin{deluxetable}{rlrrrrrlrrccr}
\rotate
\tabletypesize{\tiny \bf}
\tablecaption{Total magnitudes and model-independent parameters of
large-angular-size galaxies identified in the test fields. \label{tbl-3}}
\tablewidth{0pt}
\tablehead{
\colhead{\#} &\colhead{SDSS name} & \colhead{$u^*$} & \colhead{$g^*$} & \colhead{$r^*$} &
\colhead{$i^*$} & \colhead{$z^*$} & \colhead{b/a} & \colhead{PA}
& \colhead{R$_{\rm eff}(r)$} & \colhead{$\mu_{\rm eff}(r)$} &
\colhead{C(r)} & \colhead{a$_{26.2}(r)$} \\
& & \colhead{mag} & \colhead{mag} & \colhead{mag} &
\colhead{mag} & \colhead{mag} & &\colhead{deg} & \colhead{$^{\prime\prime}$} & \colhead{mag/$\square^{\prime\prime}$}  &
& \colhead{$^{\prime\prime}$} \\
& \colhead{(1)} & \colhead{(2)}  & \colhead{(3)}  & \colhead{(4)} &
  \colhead{(5)} & \colhead{(6)}  & \colhead{(7)}  & \colhead{(8)} &
  \colhead{(9)} & \colhead{(10)} & \colhead{(11)} & \colhead{(12)}
}
\startdata
\qq&SDSS J095146$+$001950& 18.09$\pm$  0.10& 16.59$\pm$  0.03& 15.81$\pm$  0.03& 15.42$\pm$  0.02&  15.04$\pm$  0.05& 0.47&   24&  3.6& 20.6& 3.1&  16.4    \\
\qq&SDSS J095138$+$002211& 17.44$\pm$  0.08& 16.97$\pm$  0.04& 16.60$\pm$  0.04& 16.38$\pm$  0.04&  16.47$\pm$  0.17& 0.77&$-$58& 12.3& 24.0& 2.1&  24.2    \\
\qq&SDSS J095938$+$003047& 17.31$\pm$  0.06& 16.33$\pm$  0.03& 15.87$\pm$  0.03& 15.41$\pm$  0.02&  15.17$\pm$  0.04& 0.94&   13&  3.1& 20.3& 2.3&  12.5    \\
\qq&SDSS J095939$+$003227& 16.76$\pm$  0.05& 15.70$\pm$  0.02& 15.19$\pm$  0.02& 14.83$\pm$  0.02&  14.63$\pm$  0.03& 0.46&   65&  6.9& 21.4& 3.3&  26.1    \\
\qq&SDSS J095939$+$003511& 16.68$\pm$  0.05& 15.45$\pm$  0.02& 14.74$\pm$  0.02& 14.28$\pm$  0.01&  13.97$\pm$  0.02& 0.87&   87&  7.9& 21.2& 1.8&  23.0    \\
\qq&SDSS J095959$+$003816& 17.26$\pm$  0.06& 16.55$\pm$  0.03& 16.21$\pm$  0.03& 15.89$\pm$  0.03&  15.71$\pm$  0.06& 0.92&   86&  5.3& 21.8& 2.3&  18.4    \\
\qq&SDSS J095930$-$003745& 17.97$\pm$  0.08& 17.01$\pm$  0.04& 16.45$\pm$  0.03& 16.07$\pm$  0.04&  15.45$\pm$  0.05& 0.80&$-$55&  3.4& 21.1& 2.7&  13.9    \\
\qq&SDSS J100439$+$001653& 17.91$\pm$  0.09& 16.61$\pm$  0.03& 15.97$\pm$  0.03& 15.49$\pm$  0.03&  15.30$\pm$  0.06& 0.93&$-$80&  4.3& 21.1& 3.0&  18.0    \\
\enddata
\end{deluxetable}

\begin{deluxetable}{lrrrrrrrrrrrr}
\rotate
\tabletypesize{\tiny \bf}
\tablecaption{Model-dependent photometric parameters of
large-angular-size galaxies identified in the test fields. \label{tbl-4}}
\tablewidth{0pt}
\tablehead{
\colhead{SDSS name} & \colhead{$\mu_0^d(g^*)$} & \colhead{$\alpha^d(g^*)$} &
		      \colhead{$\mu_0^b(g^*)$} & \colhead{$\alpha^b(g^*)$} &
		      \colhead{$\mu_0^d(r^*)$} & \colhead{$\alpha^d(r^*)$} &
		      \colhead{$\mu_0^b(r^*)$} & \colhead{$\alpha^b(r^*)$} &
		      \colhead{$\mu_0^d(i^*)$} & \colhead{$\alpha^d(i^*)$} &
		      \colhead{$\mu_0^b(i^*)$} & \colhead{$\alpha^b(i^*)$} \\
		    & \colhead{mag/$\square^{\prime\prime}$} & \colhead{$^{\prime\prime}$} &
		      \colhead{mag/$\square^{\prime\prime}$} & \colhead{$^{\prime\prime}$} &
		      \colhead{mag/$\square^{\prime\prime}$} & \colhead{$^{\prime\prime}$} &
		      \colhead{mag/$\square^{\prime\prime}$} & \colhead{$^{\prime\prime}$} &
		      \colhead{mag/$\square^{\prime\prime}$} & \colhead{$^{\prime\prime}$} &
		      \colhead{mag/$\square^{\prime\prime}$} & \colhead{$^{\prime\prime}$} \\
  \colhead{(1)}  & \colhead{(2)}  & \colhead{(3)}  & \colhead{(4)} & \colhead{(5)} &
  \colhead{(6)}  & \colhead{(7)}  & \colhead{(8)}  & \colhead{(9)} &
  \colhead{(10)} & \colhead{(11)} & \colhead{(12)} & \colhead{(13)}
}
\startdata
 J095146$+$001950& 21.34$\pm$0.13&   3.43$\pm$0.15& 19.61$\pm$0.14& 0.91$\pm$0.09&  20.78$\pm$0.12&   3.56$\pm$0.15& 18.52$\pm$0.09&   0.95$\pm$0.06&  20.24$\pm$0.13&   3.36$\pm$0.15& 17.98$\pm$0.10&   0.90$\pm$0.06    \\
 J095138$+$002211& 23.28$\pm$0.06&   7.64$\pm$0.22& 23.47$\pm$0.75& 1.07$\pm$0.53&  23.21$\pm$0.08&   8.54$\pm$0.30& 23.02$\pm$0.26&   1.96$\pm$0.46&  23.06$\pm$0.13&   8.84$\pm$0.51& 22.50$\pm$0.28&   1.95$\pm$0.48    \\
 J095938$+$003047& 20.42$\pm$0.08&   2.44$\pm$0.05&               &              &  19.73$\pm$0.07&   2.20$\pm$0.04&               &                &  19.29$\pm$0.10&   2.13$\pm$0.06&               &                    \\
 J095939$+$003227& 23.24$\pm$0.37&   8.39$\pm$0.78& 20.35$\pm$0.03& 1.99$\pm$0.09&  22.68$\pm$0.39&   8.07$\pm$0.78& 19.48$\pm$0.02&   1.87$\pm$0.07&  21.82$\pm$0.27&   7.47$\pm$0.47& 18.98$\pm$0.02&   1.67$\pm$0.07    \\
 J095939$+$003511& 22.36$\pm$0.29&   5.22$\pm$0.36&               &              &  21.75$\pm$0.27&   5.18$\pm$0.34&               &                &  21.05$\pm$0.15&   5.79$\pm$0.22&               &                    \\
 J095959$+$003816& 22.06$\pm$0.23&   3.51$\pm$0.21&               &              &  21.68$\pm$0.25&   3.87$\pm$0.28&               &                &  21.47$\pm$0.34&   3.80$\pm$0.37&               &                    \\
 J095930$-$003745& 21.11$\pm$0.05&   2.64$\pm$0.05&               &              &  20.41$\pm$0.06&   2.48$\pm$0.06&               &                &  19.99$\pm$0.07&   2.44$\pm$0.06&               &                    \\
 J100439$+$001653& 22.26$\pm$0.15&   5.05$\pm$0.29& 19.78$\pm$0.13& 1.13$\pm$0.09&  20.64$\pm$0.07&   3.29$\pm$0.09& 18.68$\pm$0.10&   0.75$\pm$0.04&  21.14$\pm$0.15&   5.28$\pm$0.31& 18.56$\pm$0.15&   1.07$\pm$0.10    \\
\enddata
\end{deluxetable}

\begin{deluxetable}{lrrrrrrrrr}
\rotate
\tabletypesize{\small \bf}
\tablecaption{Model-dependent photometric parameters of
large-angular-size galaxies with S\'ersic profiles
identified in the test fields.
\label{tbl-5}}
\tablewidth{0pt}
\tablehead{
\colhead{SDSS name} & \colhead{$\mu_0(g^*)$} & \colhead{$\alpha(g^*)$} & \colhead{$n(g^*)$}&
		      \colhead{$\mu_0(r^*)$} & \colhead{$\alpha(r^*)$} & \colhead{$n(r^*)$}&
		      \colhead{$\mu_0(i^*)$} & \colhead{$\alpha(i^*)$} & \colhead{$n(i^*)$}   \\
		    & \colhead{mag/$\square^{\prime\prime}$} & \colhead{$^{\prime\prime}$} & &
		      \colhead{mag/$\square^{\prime\prime}$} & \colhead{$^{\prime\prime}$} & &
		      \colhead{mag/$\square^{\prime\prime}$} & \colhead{$^{\prime\prime}$} &  \\
  \colhead{(1)}  & \colhead{(2)}  & \colhead{(3)}  & \colhead{(4)} & \colhead{(5)} &
  \colhead{(6)}  & \colhead{(7)}  & \colhead{(8)}  & \colhead{(9)} &
  \colhead{(10)}
}
\startdata
 J103806$-$000319& 23.86$\pm$0.10&   8.15$\pm$0.69&   1.55$\pm$0.13&  23.48$\pm$0.06&   7.96$\pm$0.40&   1.54$\pm$0.08&  23.16$\pm$0.10&   8.11$\pm$0.70&   1.60$\pm$0.15 \\
 J110440$+$000329& 23.68$\pm$0.06&  11.19$\pm$0.53&   1.38$\pm$0.07&  23.27$\pm$0.06&  12.25$\pm$0.53&   1.50$\pm$0.08&  22.89$\pm$0.07&  12.06$\pm$0.70&   1.23$\pm$0.07 \\
 J111055$+$010537& 22.07$\pm$0.05&   8.77$\pm$0.42&   1.17$\pm$0.04&  21.86$\pm$0.05&   9.22$\pm$0.41&   1.19$\pm$0.04&  21.71$\pm$0.08&   8.99$\pm$0.68&   1.14$\pm$0.06 \\
 J111247$-$003406& 23.93$\pm$0.07&  11.64$\pm$0.54&   1.74$\pm$0.11&  23.52$\pm$0.10&  10.12$\pm$0.70&   1.63$\pm$0.17&  23.21$\pm$0.16&   9.51$\pm$1.12&   1.51$\pm$0.22 \\
 J121257$+$011348& 24.65$\pm$0.09&   9.63$\pm$0.66&   1.43$\pm$0.11&  23.98$\pm$0.11&   8.43$\pm$0.81&   1.24$\pm$0.11&  23.68$\pm$0.13&   9.90$\pm$1.18&   1.20$\pm$0.15 \\
 J122430$+$000416& 23.73$\pm$0.06&  12.35$\pm$0.51&   1.78$\pm$0.11&  23.44$\pm$0.08&  12.47$\pm$0.65&   1.72$\pm$0.13&  23.24$\pm$0.07&  11.97$\pm$0.65&   1.55$\pm$0.11 \\
 J122622$-$011520& 23.13$\pm$0.12&  12.39$\pm$0.78&   1.62$\pm$0.12&  22.92$\pm$0.10&  11.97$\pm$0.61&   1.62$\pm$0.09&  23.16$\pm$0.13&  14.10$\pm$0.88&   1.92$\pm$0.18 \\
 J123116$+$005939& 24.24$\pm$0.08&  10.72$\pm$0.61&   1.51$\pm$0.11&  24.02$\pm$0.09&  11.23$\pm$0.68&   1.76$\pm$0.16&  23.72$\pm$0.08&  16.43$\pm$0.83&   1.70$\pm$0.19 \\
\enddata
\end{deluxetable}

\begin{deluxetable}{clllll}
\tablecaption{The Photometric Repeatability. \label{tbl:repeat}}
\tablewidth{0pt}
\tablehead{
\colhead{SDSS Name} & \colhead{$u^*$}& \colhead{$g^*$} & \colhead{$r^*$} &
\colhead{$i^*$} & \colhead{$z^*$} \\
\colhead{(1)}  & \colhead{(2)} & \colhead{(3)} & \colhead{(4)} &
\colhead{(5)}  & \colhead{(6)}
}
\startdata
SDSS J113457$-$004514  &16.97$\pm$0.06 & 16.21$\pm$0.03 & 15.90$\pm$0.03 & 15.73$\pm$0.03 & 15.39$\pm$0.06 \\
SDSS J113457$-$004514  &17.16$\pm$0.06 & 16.34$\pm$0.03 & 15.99$\pm$0.03 & 15.74$\pm$0.03 & 15.46$\pm$0.06 \\ \hline
SDSS J135230$+$002504  &17.82$\pm$0.08 & 16.68$\pm$0.03 & 15.95$\pm$0.03 & 15.46$\pm$0.02 & 15.04$\pm$0.04 \\
SDSS J135230$+$002504  &17.63$\pm$0.08 & 16.57$\pm$0.03 & 15.99$\pm$0.03 & 15.57$\pm$0.03 & 15.28$\pm$0.05 \\ \hline
SDSS J144856$-$004337  &17.96$\pm$0.08 & 17.01$\pm$0.04 & 15.92$\pm$0.03 & 16.43$\pm$0.04 & 16.05$\pm$0.08 \\
SDSS J144856$-$004337  &17.89$\pm$0.08 & 16.91$\pm$0.04 & 15.94$\pm$0.03 & 16.40$\pm$0.05 & 15.85$\pm$0.07 \\
\enddata
\end{deluxetable}

\begin{deluxetable}{cclcrc}
\tablecaption{Giant LSB spiral galaxies from our sample. \label{tbl:giant_LSBs}}
\tablewidth{0pt}
\tablehead{
\colhead{SDSS Name} &\colhead{Impey Name} & \colhead{$M_{disk}(B)$}& \colhead{$\mu_{\rm 0,c}(B)$} & \colhead{$\alpha$} & \colhead{Diff.index} \\
		    &                     & \colhead{mag}          & \colhead{mag arcsec$^2$}     & \colhead{kpc}      &  \\
\colhead{(1)}       & \colhead{(2)} & \colhead{(3)} & \colhead{(4)} & \colhead{(5)} & \colhead{(6)}
}
\startdata
SDSS J095939$+$003227&              &  $-$18.41& 24.46&   7.26& 28.76 \\
SDSS J104509$+$000433&  1042$+$0020 &  $-$20.46& 22.95&   9.32& 27.80 \\
SDSS J111117$+$010914&              &  $-$18.46& 24.48&   7.52& 28.86 \\
SDSS J120148$-$010802&              &  $-$19.87& 24.10&  12.07& 29.51 \\
SDSS J121526$-$005622&  1212$-$0039 &  $-$18.67& 24.66&   9.00& 29.43 \\
SDSS J122912$+$004903&  1226$+$0105 &  $-$21.00& 22.94&  11.88& 28.31 \\
SDSS J131821$+$002440&              &  $-$18.51& 24.69&   8.44& 29.32 \\
SDSS J132345$+$010231&  1321$+$0118 &  $-$20.54& 23.22&  10.94& 28.41 \\
SDSS J135230$+$002504&  1349$+$0039 &  $-$20.87& 22.75&  10.25& 27.80 \\
SDSS J135943$-$003424&              &  $-$20.44& 23.59&  12.37& 29.05 \\
SDSS J140831$-$000737&  1405$+$0006 &  $-$18.18& 24.59&   6.94& 28.80
\enddata
\end{deluxetable}

\end{document}